\documentclass[3p,times,twocolumn]{elsarticle}
 \biboptions{comma,sort&compress}

\usepackage{graphicx}
\usepackage{here}
\usepackage{ecrc}

\volume{00}

\firstpage{1}

\journalname{Nuclear and Particle Physics Proceedings}

\runauth{}

\jid{nppp}

\jnltitlelogo{Nuclear and Particle Physics Proceedings}

\usepackage{amssymb}

\usepackage[figuresright]{rotating}

\newcommand{\fbinv}{\mathrm{fb}^{-1}}
\newcommand{\usedLumi}{35.9~\fbinv}
\newcommand{\usedNewLumi}{41.5~\fbinv}

\newcommand{\GeV}{\mathrm{{GeV}}}
\newcommand{\TeV}{\mathrm{{TeV}}}

\newcommand{\PW}{\ensuremath{{W}}}
\newcommand{\cPZ}{\ensuremath{{Z}}}
\newcommand{\PH}{\ensuremath{{H}}}
\newcommand{\WH}{\ensuremath{\PW\PH}}
\newcommand{\ZH}{\ensuremath{\cPZ\PH}}
\newcommand{\VH}{\ensuremath{V\PH}}
\newcommand{\cPqt}{\ensuremath{{t}}}
\newcommand{\ttH}{\ensuremath{\cPqt\bar{\cPqt}\PH}}

\newcommand{\Pe}{\ensuremath{{e}}}

\newcommand{\Pgam}{\ensuremath{{\gamma}}}

\newcommand{\mgg}{\mathrm{m_{\Pgam\Pgam}}}
\newcommand{\mH}{\mathrm{m_{\PH}}}

\newcommand{\Pgm}{\ensuremath{{\mu}}}
\newcommand{\Pgmp}{\ensuremath{{\mu}^{+}}}
\newcommand{\Pgmm}{\ensuremath{{\mu}^{-}}}
\newcommand{\Pg}{\ensuremath{{g}}}
\newcommand{\HZZfl}{\ensuremath{\PH\to\cPZ\cPZ\to4\ell}}
\newcommand{\DMeVbfjj}{\ensuremath{{\mathcal D}_{\rm 2jet}}}
\newcommand{\DMeVbfj}{\ensuremath{{\mathcal D}_{\rm 1jet}}}
\newcommand{\DMeWh}{\ensuremath{{\mathcal D}_{\rm \WH}}}
\newcommand{\DMeZh}{\ensuremath{{\mathcal D}_{\rm \ZH}}}

\newcommand{\ttbar}{\ensuremath{\cPqt\bar{\cPqt}}}
\newcommand{\KD}{\ensuremath{{\cal D}^{\rm kin}_{\rm bkg}} }

\newcommand{\MassDprime}{\mathrm{{\cal D}'_{\rm mass}} }

\newcommand{\pt}{\ensuremath{p_{\mathrm{T}}}}

\newcommand{\mZ}{\ensuremath{m_{Z}}}

\newcommand{\valMassThreeDRefit}{\ensuremath{125.26 \pm 0.20 (\mathrm{stat.}) \pm 0.08 (\mathrm{sys.})}}

\newcommand{\zero}{0-}
\newcommand{\twol}{2-lepton}
\newcommand{\one}{1-}

\newcommand{\bbbar}{\ensuremath{b\bar{b}}}

\newcommand{\HBB}{\ensuremath{\PH\to\bbbar}}

\newcommand{\mmm}{\ensuremath{m_{\Pgm\Pgm}}}
\newcommand{\Htomm}{\ensuremath{\PH\!\to\!\Pgmp\Pgmm}}

\newcommand{\stat}{\ensuremath{(\mathrm{stat.})}}
\newcommand{\syst}{\ensuremath{(\mathrm{sys.})}}
\newcommand{\thy}{\ensuremath{(\mathrm{theo.})}}

\newcommand{\Pgt}{\ensuremath{\mathrm{\tau}}}

\newcommand{\sigmod}{\ensuremath{\mu_{t\bar{t}\PH}}}

\newcommand{\CL}{\ensuremath{CL}}

\def\Acknowledgements{\bigskip  \bigskip \begin{center} \begin{large}
             \bf ACKNOWLEDGEMENTS \end{large}\end{center}}

\begin{document}

\begin{frontmatter}

\title{ Measurements of the 125~$\GeV$ Higgs boson at CMS
 $^*$}
 \cortext[cor0]{Talk given at 21th International Conference in Quantum Chromodynamics (QCD 18),  2 July - 6 July 2018, Montpellier - FR}
 \author[label1]{Junquan Tao}
   \ead{taojq@mail.ihep.ac.cn}

\address[label1]{Institute of High Energy Physics, Chinese Academy of Sciences, Beijing 100049, P. R. China}
 \author[]{on behalf of the CMS Collaboration}

\pagestyle{myheadings}
\markright{ }
\begin{abstract}
Results of the measurements of the 125~$\GeV$ Higgs boson properties with proton-proton collision data at $\sqrt{s}=13~\TeV$ collected by CMS detector are presented.
The used Higgs boson decay channels include the five major decay modes, $\mathrm{H}\rightarrow\gamma\gamma$, $\mathrm{H}\rightarrow{\rm Z}{\rm Z}\rightarrow4\ell$, $\mathrm{H}\rightarrow{\rm W}{\rm W}\rightarrow\ell\nu\ell\nu$, $\mathrm{H}\rightarrow\tau^{+}\tau^{-}$ and $\mathrm{H}\rightarrow b\bar{b}$, and two rare decay modes, $\mathrm{H}\rightarrow\mu^{+}\mu^{-}$ and $\mathrm{H}\rightarrow{\rm Z}/\gamma^{*}+\gamma\rightarrow\ell\ell\gamma$, with $\ell={\rm e},\mu$.
The measured Higgs boson properties include its mass, signal strength relative to the standard model prediction, signal strength modifiers for different Higgs boson production modes, coupling modifiers to fermions and bosons, effective coupling modifiers to photons and gluons, simplified template cross sections, fiducial cross sections. All results are consistent, within their uncertainties,
with the expectations for the Standard Model Higgs boson.
\end{abstract}
\begin{keyword}

Higgs boson properties \sep Standard Model

\end{keyword}

\end{frontmatter}
\section{Introduction}

The standard model of particle physics (SM)~\cite{Glashow:1961tr,Weinberg:1967tq,Salam:1968rm}
has been very successful in explaining high-energy experimental data. During the
Run~1 period of the CERN LHC, a new particle was discovered by both
ATLAS~\cite{Aad:2012tfa} and CMS~\cite{Chatrchyan:2012xdj,Chatrchyan:2013lba} experiments
and the collected experimental evidence is consistent with the particle
being a Higgs boson~\cite{Aad:2015zhl,Khachatryan:2016vau,Khachatryan:2014kca,Khachatryan:2016ctc} compatible with the quantum of the scalar field postulated by the
Higgs mechanism~\cite{Englert:1964et,Higgs:1964pj,Guralnik:1964eu}.

In this proceeding the results of the measurements of the 
Higgs boson properties at CMS~\cite{Chatrchyan:2008aa} are summarized.
For most of the results, the proton-proton (pp) collision data recorded by the CMS detector during 2016 (2016 data set), corresponding to an integrated luminosity
of $\usedLumi$, are used. For $\mathrm{H}\rightarrow{\rm Z}{\rm Z}$ and $\mathrm{H}\rightarrow b\bar{b}$, the data set corresponding to an integrated luminosity of $\usedNewLumi$ at $\sqrt{s}=13~\TeV$ and collected by CMS detector in 2017 (2017 data set) are also used.
There are four main mechanisms for Higgs boson production in pp collisions at LHC.
The gluon-gluon fusion mechanism (ggH) has the largest cross section,
followed in turn by vector boson fusion (VBF), associated production with a vector boson (VH with V $=$ $\PW$ or $\cPZ$) and production in association with top quarks, $\ttH$.
The results presented here are based
on the analyses of the following five decay modes: $\mathrm{H}\rightarrow\gamma\gamma$~\cite{CMS:2017rli,CMS:2017nyv}, $\mathrm{H}\rightarrow{\rm Z}{\rm Z}$
followed by ZZ decays to 4 leptons~\cite{Sirunyan:2017exp,Sirunyan:2017tqd,CMS:2018mmw}, $\mathrm{H}\rightarrow{\rm W}{\rm W}$ followed by ${\rm W}{\rm W}\rightarrow \ell \nu \ell \nu$ decays~\cite{Sirunyan:2018egh}, $\mathrm{H}\rightarrow\tau^{+}\tau^{-}$~\cite{Sirunyan:2017khh,Sirunyan:2018cpi} and $\mathrm{H}\rightarrow b\bar{b}$~\cite{Sirunyan:2017elk,Sirunyan:2018kst}, the following two rare decays modes: $\mathrm{H}\rightarrow Z/\gamma^{*}\gamma$ followed by $Z/\gamma^{*}$ decays to 2 leptons ($\cPZ/\gamma^{*} \rightarrow \ell  \ell$)~\cite{Sirunyan:2018tbk} and $\mathrm{H}\rightarrow\mu^{+}\mu^{-}$~\cite{Sirunyan:2018hbu}, and the combined measurements of Higgs boson couplings in pp collisions at $\sqrt{s}=13$ TeV~\cite{CMS:2018lkl}. Here and
throughout, $\ell$ stands for electrons or muons ($\ell={\rm e},\mu$).

The structure of this proceeding is organized as follows.
In section 2 the analysis strategy of each Higgs boson decay channel is introduced briefly.
Section 3 shows the results from each decay channel and also the combined measurements of the Higgs boson¡¯s production and decay rates, as well its
couplings to vector bosons and fermions.
Finally the conclusions are 
in section 4. 

\section{Analysis strategy}

\subsection{$\mathrm{H}\rightarrow\gamma\gamma$}

Despite the small branching ratio predicted by the SM ($\approx$ 0.2\%),
the $\mathrm{H}\rightarrow\gamma\gamma$ decay channel provides a clean final
state with an invariant mass peak that can be reconstructed with high
precision.
So a search is made for a fully reconstructed peak
in the diphoton invariant mass distribution
on a large irreducible background from QCD production
of two photons.
A Boosted Decision Tree (BDT) is used to separate prompt photons from
photon candidates satisfying the preselection requirements while
resulting from misidentification of jet fragments.
The diphoton vertex assignment relies on another BDT,
whose inputs are observables related to tracks
recoiling against the diphoton system.
To improve the sensitivity of the analysis, events are classified
targeting different production mechanisms and according to their mass
resolution and predicted signal-to-background ratio. A dedicated diphoton multivariate classifier
(also BDT) is trained to evaluate the diphoton mass resolution on a per-event basis and is used
as an ingredient in the categorization.
To extract signal events parametric models for signal and background have been
built separately for each category. The signal model is extracted from simulation
as a combination of several gaussian function taking into account different
correction and scale factors. The background models are completely data driven where a
nuisance parameter is set to vary over a set of possible functional forms. This
technique is discussed in details in~\cite{Dauncey:2014xga}.

\subsection{$\mathrm{H}\rightarrow ZZ $}

The $\HZZfl$ decay channel has a large signal-to-background ratio, and the precise
reconstruction of the final-state decay products allows the complete determination of the
kinematics of the Higgs boson. This makes it one of the most important channels for studies of the Higgs boson's properties.

The full kinematic information from each event using either the Higgs boson decay products or
associated particles in its production is extracted using matrix element calculations and used
to form several kinematic discriminants.
The discriminant sensitive to the $\Pg\Pg/q\bar{q}\to4\ell$ kinematics, $\mathcal{D}^{\rm kin}_{\rm bkg}$,
is calculated as~\cite{Chatrchyan:2012xdj,Khachatryan:2014kca}. Four discriminants calculated as~\cite{Khachatryan:2015cwa,Khachatryan:2015mma}
are used to enhance the purity of event categories which are sensitive to the VBF signal topology with two associated jets ($\DMeVbfjj$),
the VBF signal topology with one associated jet (\DMeVbfj), and to the VH (either ZH or WH) signal topology with two associated jets from the decay of the Z$\to{\rm q\bar{q}}$ or the W$\to{\rm q\bar{q}^\prime}$ ($\DMeWh$ and $\DMeZh$).

In order to improve the sensitivity to the Higgs boson production mechanisms,
the selected events are classified into seven categories,
based on the multiplicity of jets, b-tagged jets and additional leptons, missing energy and selections on kinematic discriminants $\mathcal{D}^{\rm kin}_{\rm bkg}$.

\subsection{$\mathrm{H}\rightarrow W^{+}W^{-} $}

The large Higgs boson branching fraction to a W boson pair
makes this channel one of the most suitable for a precision measurement of the Higgs boson
production cross section, and also allows studies of subleading production channels, such as
VBF and VH.

The leptonic decays of the two W bosons provide the cleanest decay channel, despite the presence
of neutrinos in the final state that prevents the full reconstruction of the Higgs boson mass.
The different-flavor leptonic decay mode $e\mu$ has the largest branching fraction, is the least
affected by background processes, and therefore is the most sensitive channel of the analysis.
The same-flavor $e^{+}e{-}$ and $\mu^{+}\mu^{-}$ final states are also considered, although their sensitivity
is limited by the contamination from the Drell-Yan background with missing transverse
momentum due to instrumental effects.

The $W^{+}W^{-}$ candidates are selected in events with an oppositely
charged lepton pair, large missing transverse momentum, and various numbers
of jets.
The events are categorized by jet multiplicity to better handle the $t\bar{t}$ background. In addition,
dedicated categories are designed to enhance the sensitivity to the VBF and VH production
mechanisms.

\subsection{$\mathrm{H}\rightarrow\tau^{+}\tau^{-}$}

To establish the mass generation mechanism for fermions, it is necessary to probe the direct coupling of the Higgs boson to such particles. The most promising decay channel is $\tau^{+}\tau^{-}$, because of the large event rate expected in the SM compared to the $\mu^{+}\mu^{-}$ decay channel. Using pp collision data at Run~1,
the combination of the results from CMS and ATLAS experiments yields an observed (expected) significance of 5.5 (5.0) standard deviations~\cite{Khachatryan:2016vau}.

Measurement of the $\mathrm{H}\rightarrow\tau^{+}\tau^{-}$ signal strength is performed using 2016 data set.
All possible $\tau^{+}\tau^{-}$ final states are studied, except for those with two muons or two electrons because of the low branching fraction and large background contribution. In the following,
the symbol $\tau_{h}$ refers to $\tau$ leptons reconstructed in their hadronic decays. The four $\tau$-pair final states with the largest branching fractions are studied in
this analysis : $\mu\tau_{h}$, $e\tau_{h}$, $\tau_{h}\tau_{h}$, and $e\mu$.

\subsection{$\mathrm{H}\rightarrow b\bar{b}$}

The $\mathrm{H}\rightarrow b\bar{b}$ decay tests directly the Higgs boson coupling to
fermions, and more specifically to down-type quarks, and has not yet been established
experimentally. In the SM, for a Higgs boson mass $\mH = 125$~$\GeV$, the branching fraction is approximately 58\%, by far the largest.
An observation in this channel is necessary to solidify the Higgs boson as the source of mass generation in the fermion sector of the SM~\cite{Weinberg:1967tq,Nambu:1961fr}.

A search for the $\mathrm{H}\rightarrow b\bar{b}$ when Higgs boson produced in association with an electroweak vector boson is summarized in this proceeding for the following processes: $\cPZ(\nu\nu)\PH$, $\PW(\mu\nu)\PH$, $\PW(\Pe\nu)\PH$, $\cPZ(\mu\mu)\PH$, and $\cPZ(\Pe\Pe)\PH$, using both 2016 and 2017 data set~\cite{Sirunyan:2017elk,Sirunyan:2018kst}.
The final states that
predominantly correspond to these processes, respectively, are characterized by the number of leptons required in the event selection, and are referred to as the \zero, \one, and \twol\ channels.
The results from this search are combined with those of similar searches performed by the CMS Collaboration during Run~1~\cite{Chatrchyan:2013zna,Khachatryan:2015bnx}.

\subsection{$\mathrm{H}\rightarrow Z/\gamma^{*}(\rightarrow \ell  \ell)\gamma$}

Measurements of rare decays of the Higgs boson, such as
$\PH\to\gamma^*\gamma$ and $\PH\to\cPZ\gamma$, would enhance our
understanding of the SM of particle physics, and
allow us to probe exotic couplings introduced by possible extensions
of the SM~\cite{Chen:2012ju,Sun:2013rqa,Passarino:2013nka}.

In the search for $\PH\to Z/\gamma^* + \gamma$, the leptonic channel,
$\gamma^*/\cPZ \to \ell\ell$ ($\ell=\Pe$ or $\mu$), is most promising as it
has relatively low background.
Experimentally one can separate the  off- and on-shell
contributions, and define the respective signal regions, using a selection
based on the invariant mass of the dilepton
system, $m_{\ell\ell} = m_{\gamma^*/\cPZ}$. For the measurements presented in this proceeding
a threshold of $m_{\ell\ell} = 50~\GeV$ is used to
separate the two processes.
This proceeding will summarize the results from the search for Higgs boson decaying to
 $\PH\to\gamma^*\gamma\to\mu\mu\gamma$ and
 $\PH\to\cPZ\gamma\to\ell\ell\gamma$ at 13~$\TeV$.

\subsection{$\mathrm{H}\rightarrow\mu^{+}\mu^{-}$}

The study of the Higgs boson decays
to muons is of particular importance because it extends the investigation to
its couplings to fermions of the second generation.  For a Higgs boson with a
mass of 125.09~$\GeV$, the expected branching fraction to muons is
$2.17 \times 10^{-4}$~\cite{deFlorian:2016spz}.
The signal would appear as a narrow resonance over a smoothly
falling mass spectrum from the SM background processes, primarily Drell--Yan
and leptonic $\ttbar$ decays.

This proceeding will summarize the result, the upper limits on the product of the
Higgs boson production cross section and branching fraction
$\mathcal{B}(\Htomm)$, of the search for $\Htomm$
events with 2016 data set
, and its combination with the Run~1 data collected at $\sqrt{s} = 7$ and 8~$\TeV$
corresponding to integrated luminosities of 5.0~$\fbinv$ and 19.7~$\fbinv$,
respectively.
Events are classified into categories using variables that are largely
uncorrelated with $\mmm$ in order to enhance the sensitivity to the Higgs boson
signal. The primary Higgs boson production mechanisms targeted by this analysis
are VBF and $\Pg\Pg\PH$.
The event categories are defined using the BDT score trained to distinguish between
the signal events and the background, and the expected dimuon
mass resolution, gauged by the largest $|\eta|$ of the two muons.

\subsection{Combined measurements}

Combined measurements of the production and decay rates of the Higgs boson, as well as its couplings to vector bosons and fermions, using 2016 data set are also presented.
The combination is based on the analyses targeting the four main Higgs boson production mechanisms (gluon fusion, vector boson fusion, and associated production with a W or Z boson, or a top quark-antiquark pair) and the following decay modes: H $\to$ $\gamma\gamma$, ZZ, WW, $\tau\tau$, bb, and $\mu\mu$, as mentioned above. Searches for invisible Higgs boson decays are also considered.

\section{Results}

\subsection{Mass}

\begin{figure}[htbp]
\centering
\includegraphics[width=3.8cm]{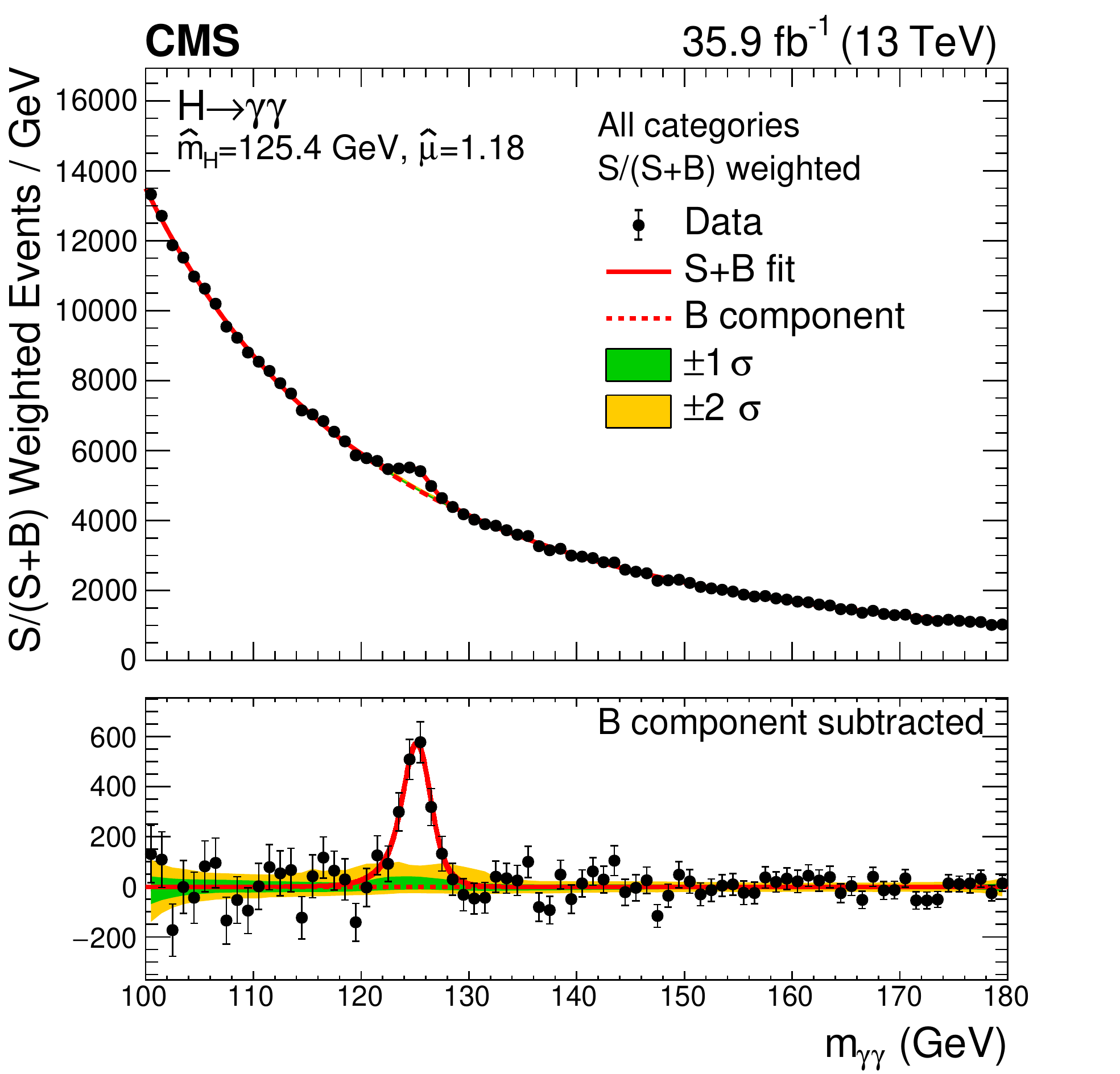}
\includegraphics[width=3.8cm]{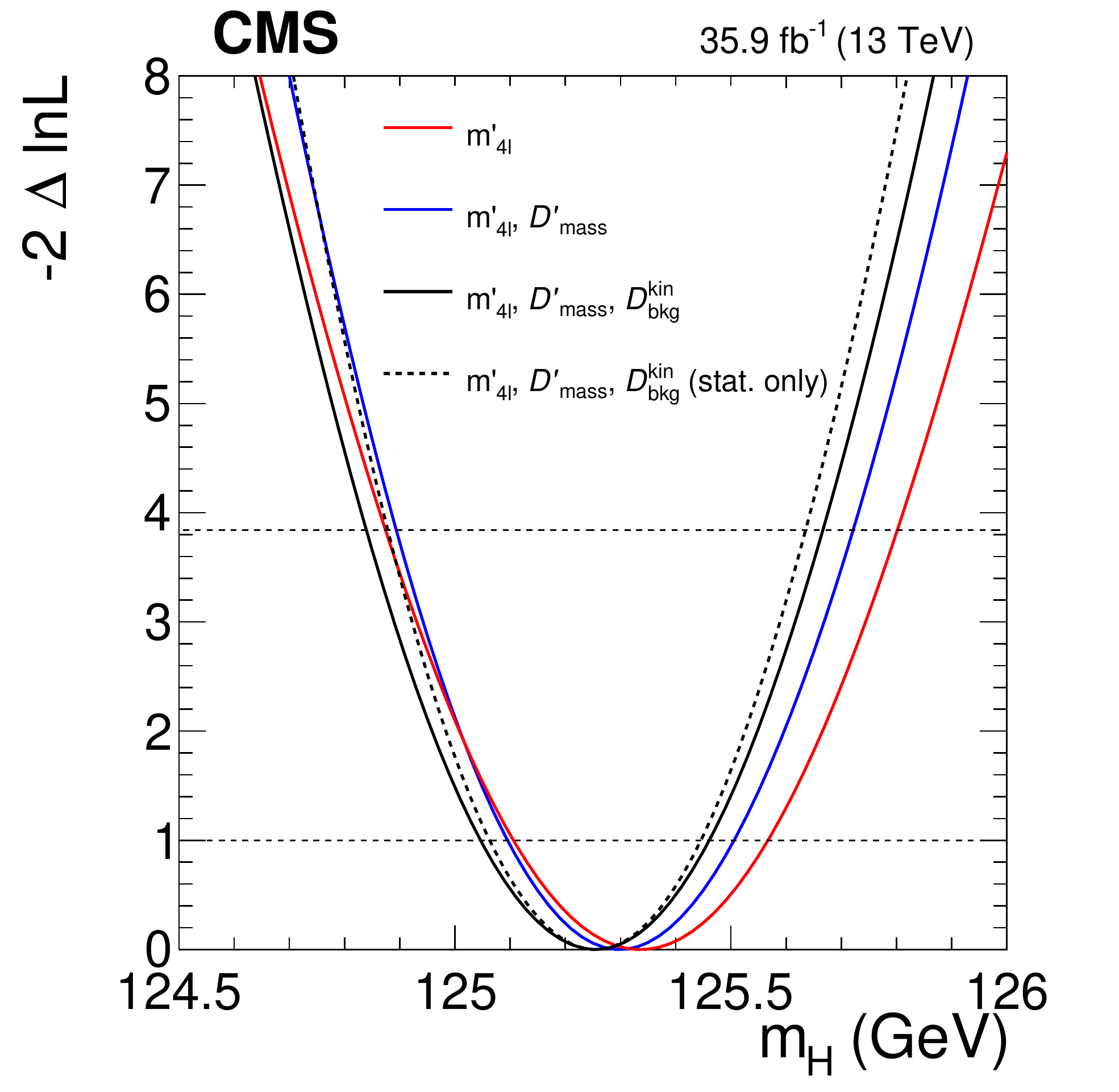}
\caption{ Left plot shows $\mgg$ distribution in data and signal-plus-background model fits
for all categories summed and
weighted by their sensitivity~\cite{CMS:2017rli}. The one (green)
and two (yellow) standard deviation bands include the uncertainties in the
background component of the
fit. The lower panel shows the residuals after the background subtraction. Right plot gives 1D likelihood scans as a function of the Higgs boson mass for the 1D, 2D, and 3D measurement from the $\HZZfl$ analysis~\cite{Sirunyan:2017exp}. The likelihood scans are shown for the mass measurement using the refitted mass distribution with the $m(\cPZ_1)$ constraint.
      Solid lines represent scans with all uncertainties included, dashed lines those with only statistical uncertainties.
} \label{fig:HggHZZMass}
\end{figure}

The left plot in Figure~\ref{fig:HggHZZMass} shows the $\mgg$ distribution in data and signal-plus-background model fits
for all categories summed and
weighted by their sensitivity. The best fit mass is found at $m_{\PH}$
$= 125.4\pm 0.3~\GeV = 125.4
\pm 0.2\stat \pm 0.2\syst~\GeV$, compatible with the combined mass
measurement from ATLAS and CMS~\cite{Aad:2015zhl}.
In the analysis of $\HZZfl$, different distributions are used to build the likelihood used to extract the Higgs boson mass.
The one dimensional (1D) likelihood scans vs. $\mH$, while profiling the signal strength modifier
$\mu$ along with all other nuisance parameters for the 1D $\mathcal{L}(m_{4\ell}')$, 2D $\mathcal{L}(m_{4\ell}',\MassDprime)$,
and 3D $\mathcal{L}(m_{4\ell}',\MassDprime,\KD)$ fits, including the $m(\cPZ_1)$ constraint, with $\cPZ_1$ the $\cPZ$ candidate with an invariant
mass closest to the nominal $\cPZ$ boson mass ($\mZ$)~\cite{Patrignani:2016xqp},
are shown in the right of Figure~\ref{fig:HggHZZMass}. The nominal result for the mass measurement is obtained from the 3D fit. 
The measured mass value from $\HZZfl$ is $m_{\PH} = \valMassThreeDRefit~\GeV$.
The mass measurement from $\HZZfl$ is the most precise at LHC. It is statistical limited while them main systematic uncertainty is the lepton momentum scale.

\subsection{Signal strengths}

\begin{figure}[htbp]
\centering
\includegraphics[width=4.2cm]{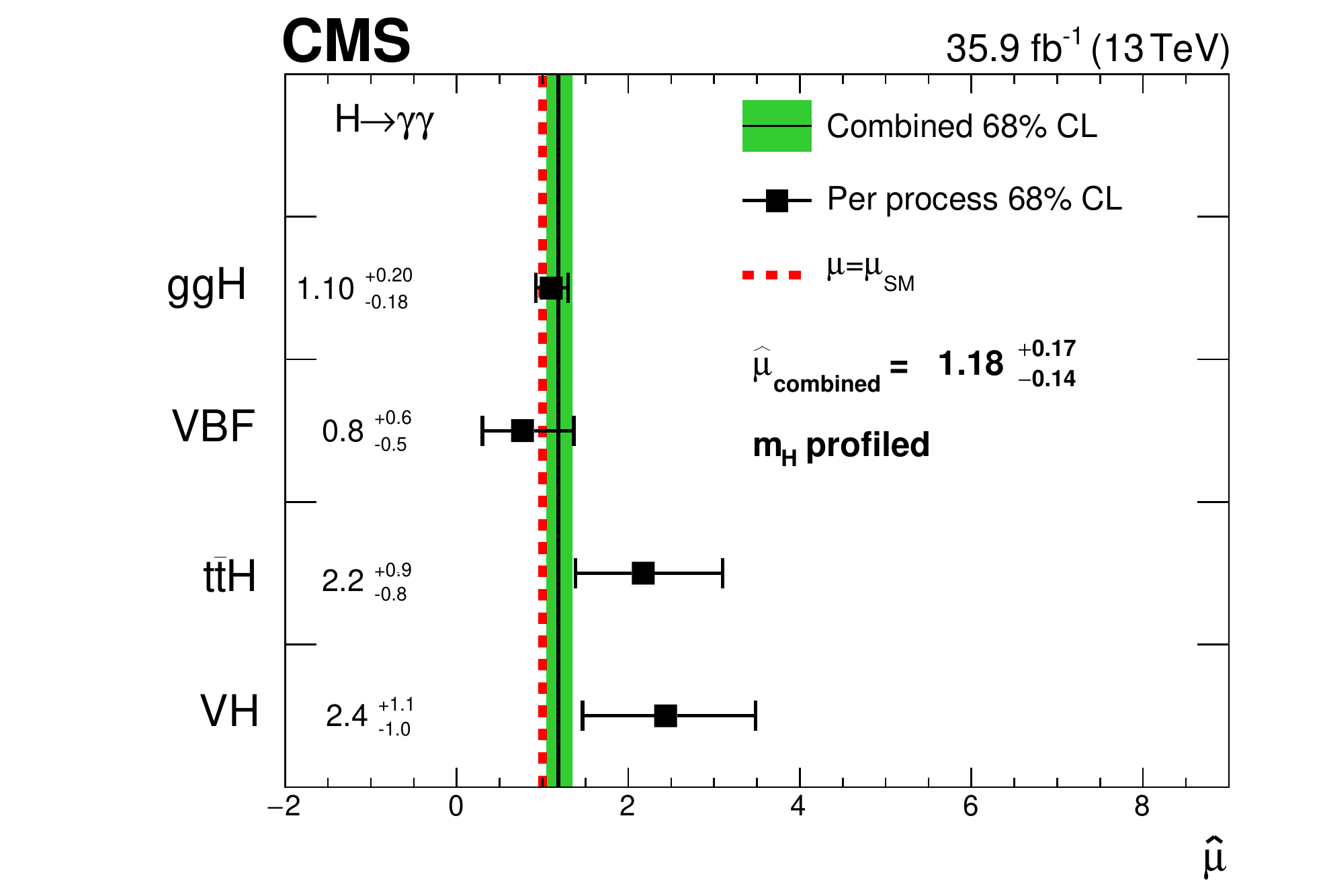}
\includegraphics[width=3.4cm]{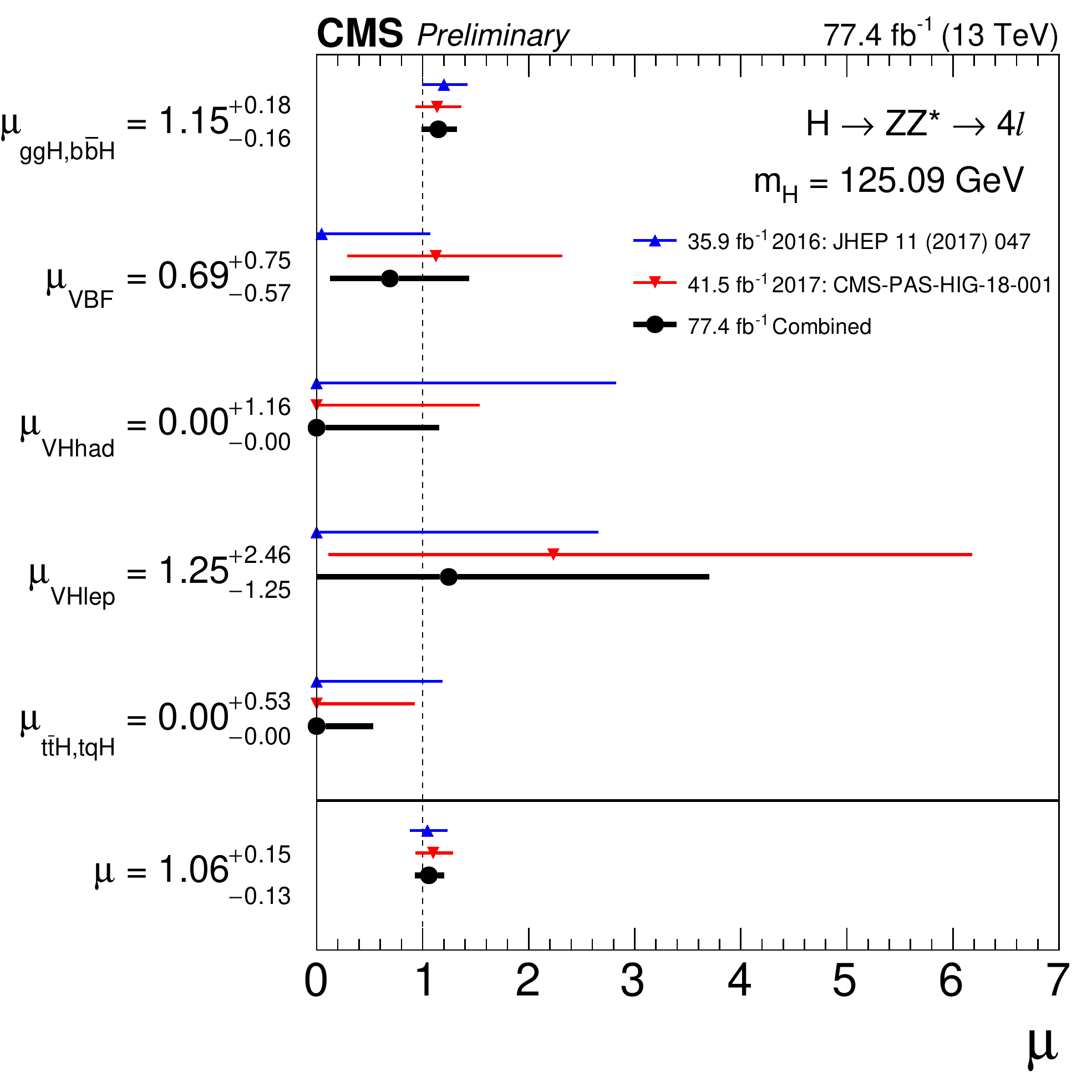}
\caption{ The measured signal strength modifier for each production mode and the global signal strength at the best fit mass value $m_{\PH}$
$= 125.4~\GeV$, with $\mathrm{H}\rightarrow\gamma\gamma$ channel (left)~\cite{CMS:2017rli}. Signal strength modifiers corresponding to the main SM Higgs boson
production modes and the global signal strength at $m_{\mathrm{H}}=125.09~\mathrm{GeV}$, in each 2016 and 2017 data and their combination, with $\mathrm{H}\rightarrow{\rm Z}{\rm Z}\rightarrow4\ell$ decay channel (right)~\cite{CMS:2018mmw}.
} \label{fig:HSigStrength}
\end{figure}

A likelihood scan of the signal strength modifier, defined
as the ratio of the observed Higgs boson rate in each decay channel to
the standard model expectation, is performed.
From $\mathrm{H}\rightarrow\gamma\gamma$ decay channel, the best fit signal strength modifier measured for all categories
combined using this method is
$\mu = 1.18^{+0.17}_{-0.14}=1.18\ ^{+0.12}_{-0.11}\stat ^{+0.09}_{-0.07}\syst ^{+0.07}_{-0.06}\thy$ at the best fit mass value $m_{\PH}$
$= 125.4~\GeV$ as mentioned above. The results of a fit to the signal strength modifier for each production
mode, defined analogously to the overall $\mu$ above, are shown in the left of
Figure~\ref{fig:HSigStrength}. A simultaneous fit to all categories is performed to extract the signal strength modifier for the $\mathrm{H}\rightarrow{\rm Z}{\rm Z}\rightarrow4\ell$ decay channel. With the 2017 data set, 
the combined signal strength is measured to be $\mu = 1.10^{+0.19}_{-0.17} = 1.10^{+0.14}_{-0.13} ({\rm stat}) ^{+0.13}_{-0.14} ({\rm syst})$ at the
combined measured mass from ATLAS and CMS in Run~1 $m_{\mathrm{H}}=125.09~\mathrm{GeV}$. The measured signal strength modifier
from the combination of 2016 and 2017 data is $\mu = 1.06^{+0.15}_{-0.13}$.
The right plot of Figure~\ref{fig:HSigStrength} shows
the signal strength modifiers corresponding to the main SM \\
Higgs boson production modes and the global signal strength $\mu$,
in each 2016 and 2017 data and their combination. 
SM expectation is 1.
From $\mathrm{H}\rightarrow W^{+}W^{-}$ decay channel, the observed cross section times branching fraction is $1.28 ^{+0.18}_{-0.17}$ times the standard model prediction for the Higgs boson with a mass of $125.09\GeV$.

\subsection{Cross section measurements}

Various measurements of the cross section for Higgs boson production are performed. Firstly the cross section measurements for different SM Higgs boson production
processes ($\sigma_{\Pg\Pg\PH}$, $\sigma_{\mathrm{VBF}}$, $\sigma_{\rm VHhad}$, $\sigma_{\rm VHlep}$, and $\sigma_{\ttH}$) in the simplified template cross section (STXS) framework~\cite{deFlorian:2016spz} for a reduced fiducial volume defined using a selection on the Higgs boson rapidity $|y_{\rm H}|<2.5$. The STXS approach differs from the signal strength modifier measurements in the splitting of the production modes, and reduces the dependence of the measurements on the theoretical uncertainties in the SM predictions, by avoiding the size able uncertainty associated with the extrapolation to the full phase space. The measurements correspond
to the `stage-0' simplified template cross sections from~\cite{deFlorian:2016spz}. The left plot of Figure~\ref{fig:HSigStrength} gives the measured cross section ratios from the $\mathrm{H}\rightarrow\gamma\gamma$ channel for each process (black points) in the Higgs boson simplified
template cross section framework, with the SM Higgs boson mass profiled, compared to
the SM expectations and their uncertainties (blue band). The right plot of Figure~\ref{fig:HSigStrength} shows the measured simplified template cross sections, normalized
to the SM prediction, from $\mathrm{H}\rightarrow{\rm Z}{\rm Z}\rightarrow4\ell$ decay channel. The grey bands indicate the theoretical uncertainties in the SM
predictions. The orange error bars show the full uncertainty, including experimental uncertainties
and theoretical uncertainties causing migration of events between the various categories.

\begin{figure}[htbp]
\centering
\includegraphics[width=4.2cm]{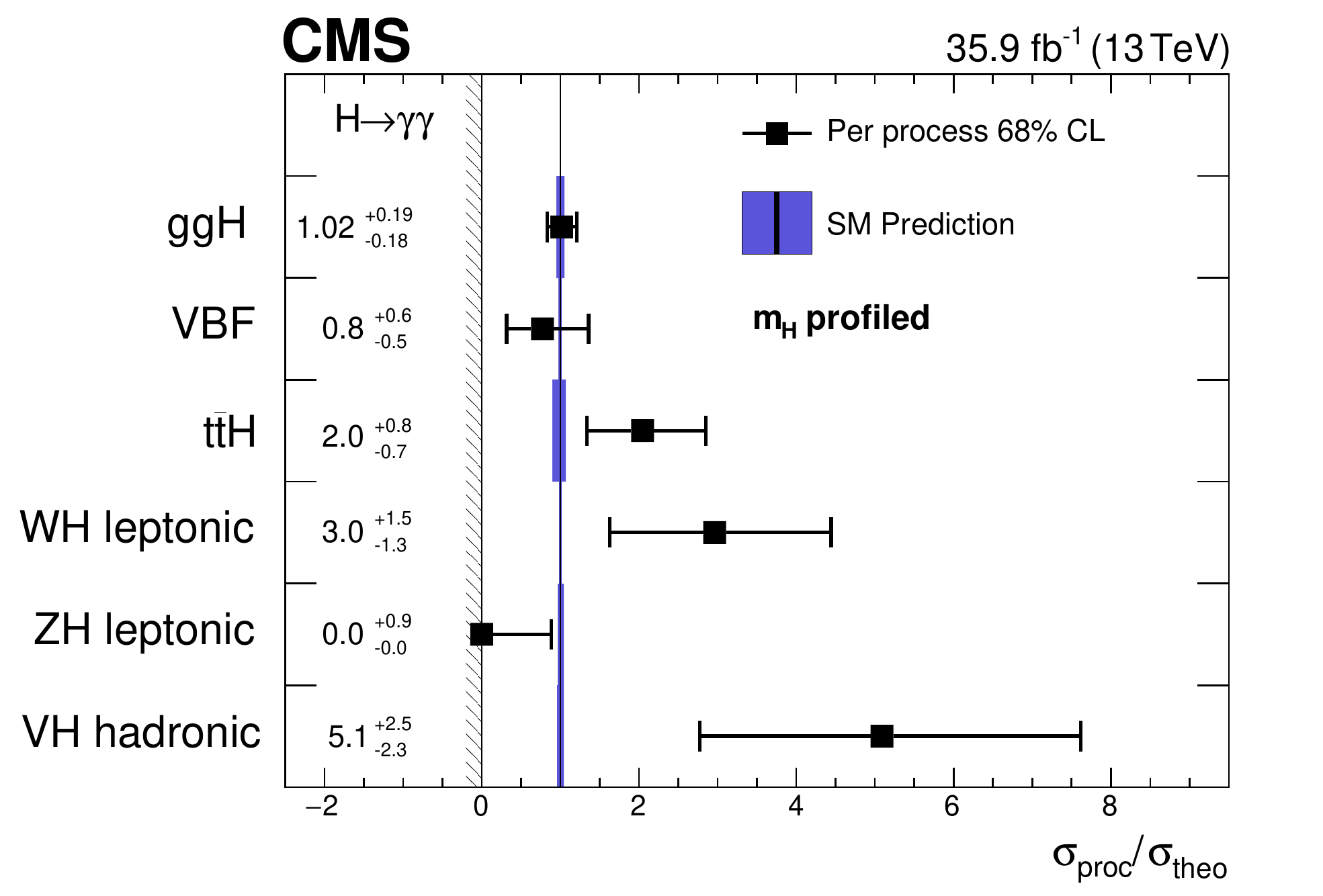}
\includegraphics[width=3.4cm]{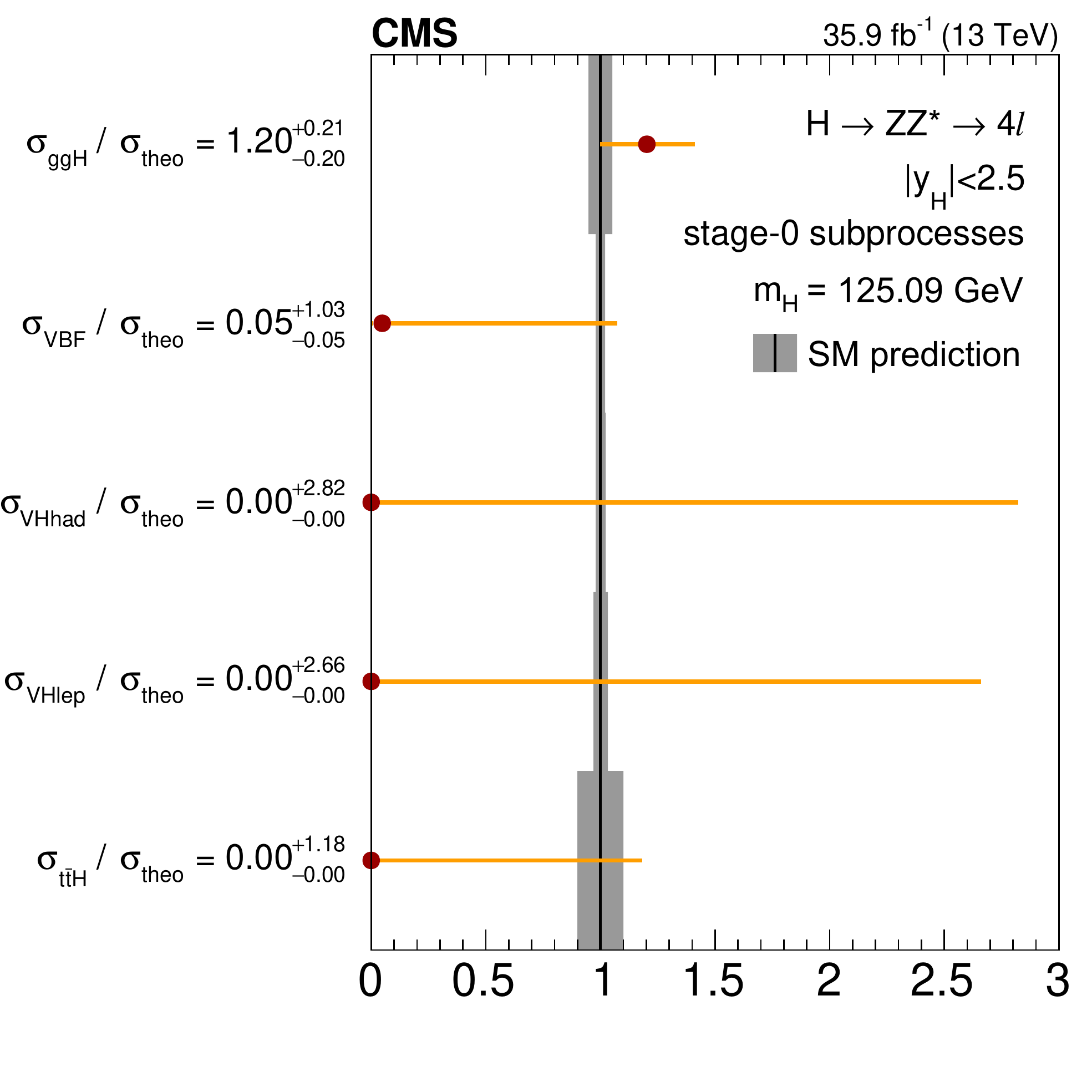}
\caption{ The measured cross section ratios from the $\mathrm{H}\rightarrow\gamma\gamma$ channel for each process (black points) in the Higgs boson simplified
template cross section framework, with the SM Higgs boson mass profiled, compared to
the SM expectations and their uncertainties in blue band (left plot)~\cite{CMS:2017rli}. The measured simplified template cross sections, normalized
to the SM prediction, from $\mathrm{H}\rightarrow{\rm Z}{\rm Z}\rightarrow4\ell$ decay channel (right plot)~\cite{Sirunyan:2017exp}. The grey bands indicate the theoretical uncertainties in the SM
predictions. The orange error bars show the full uncertainty, including experimental uncertainties
and theoretical uncertainties causing migration of events between the various categories.
} \label{fig:HiggsSTXS}
\end{figure}

\begin{figure}[htbp]
\centering
\includegraphics[width=3.7cm]{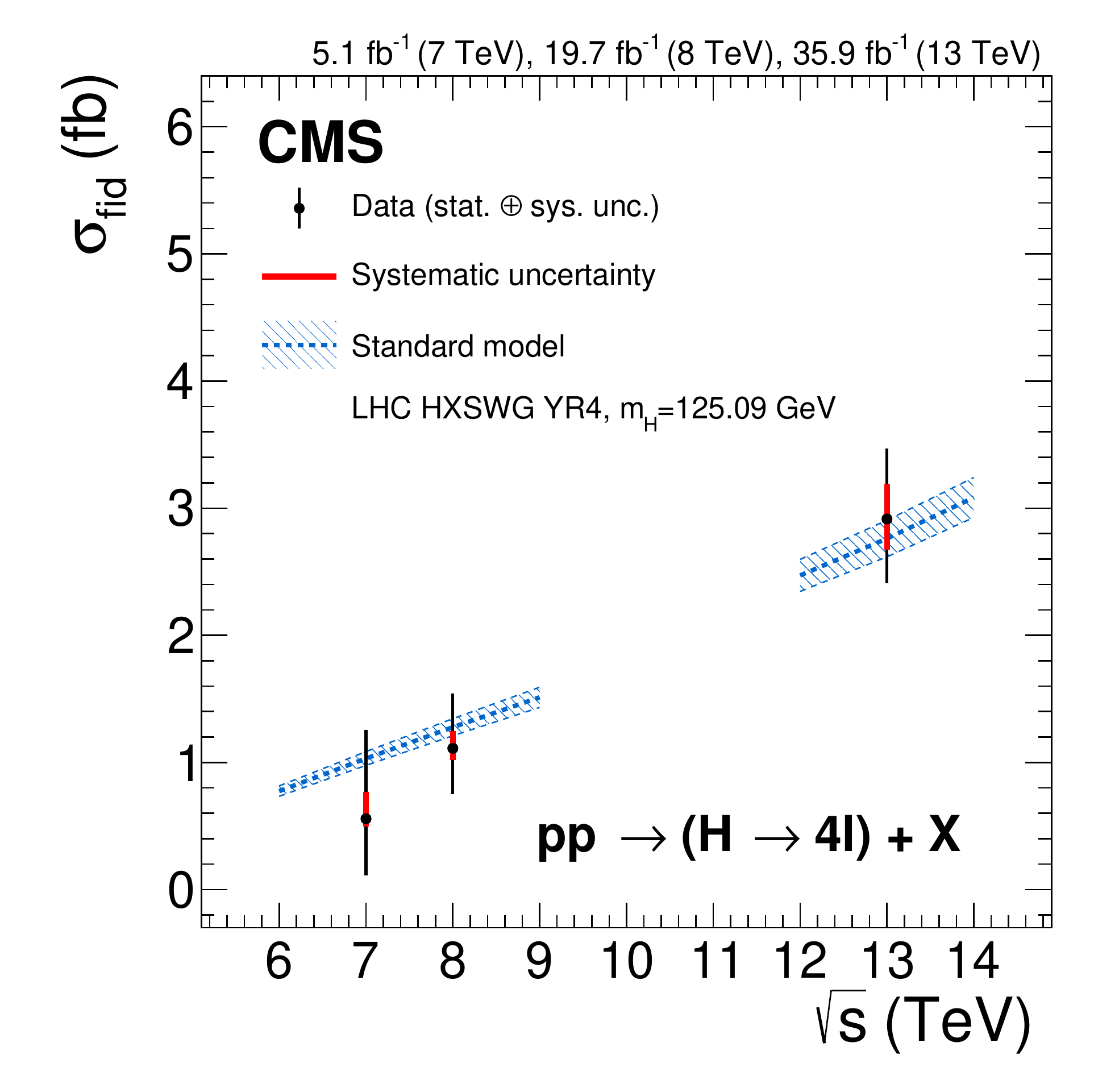}
\includegraphics[width=3.7cm]{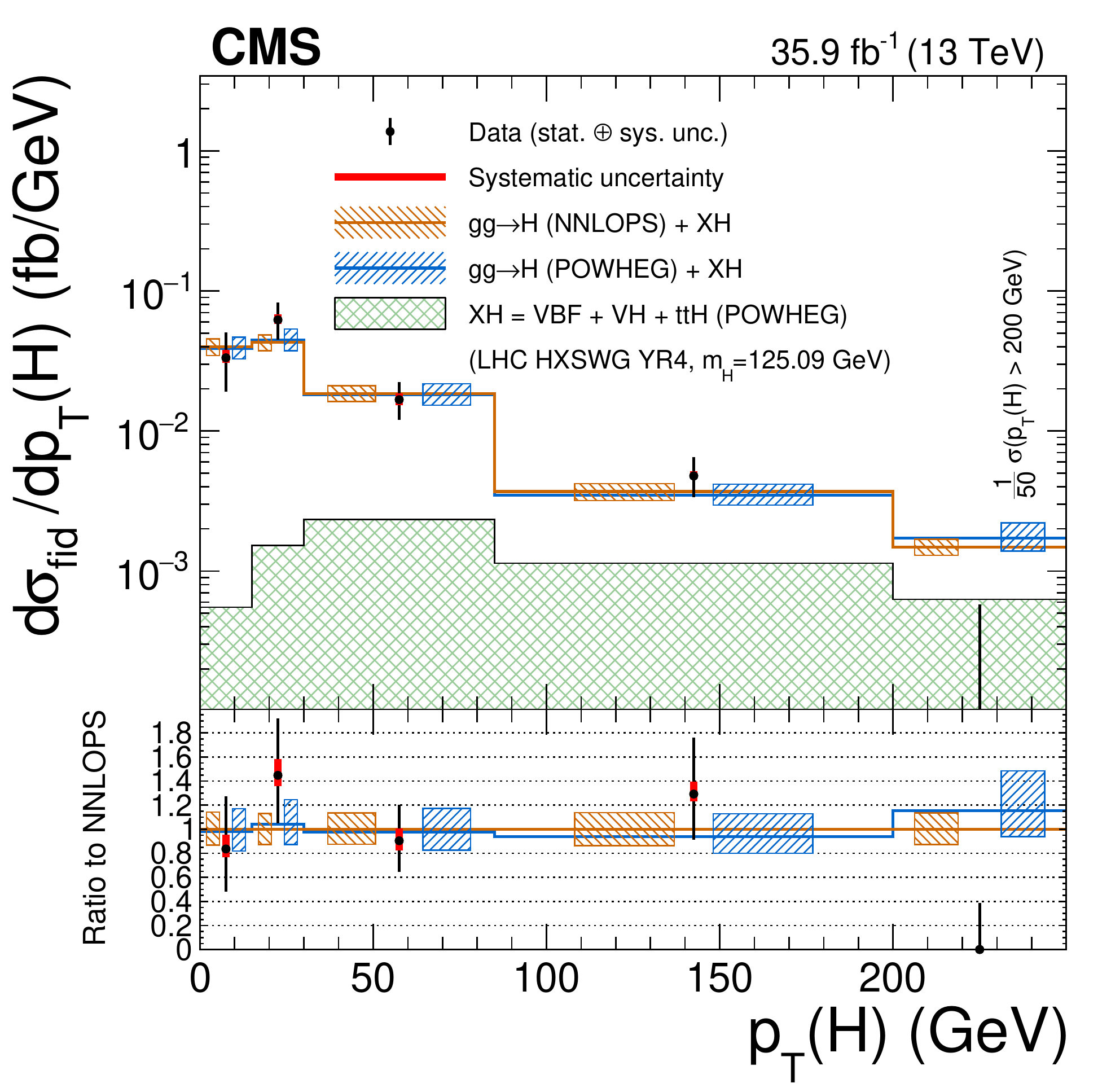} \\
\includegraphics[width=3.7cm]{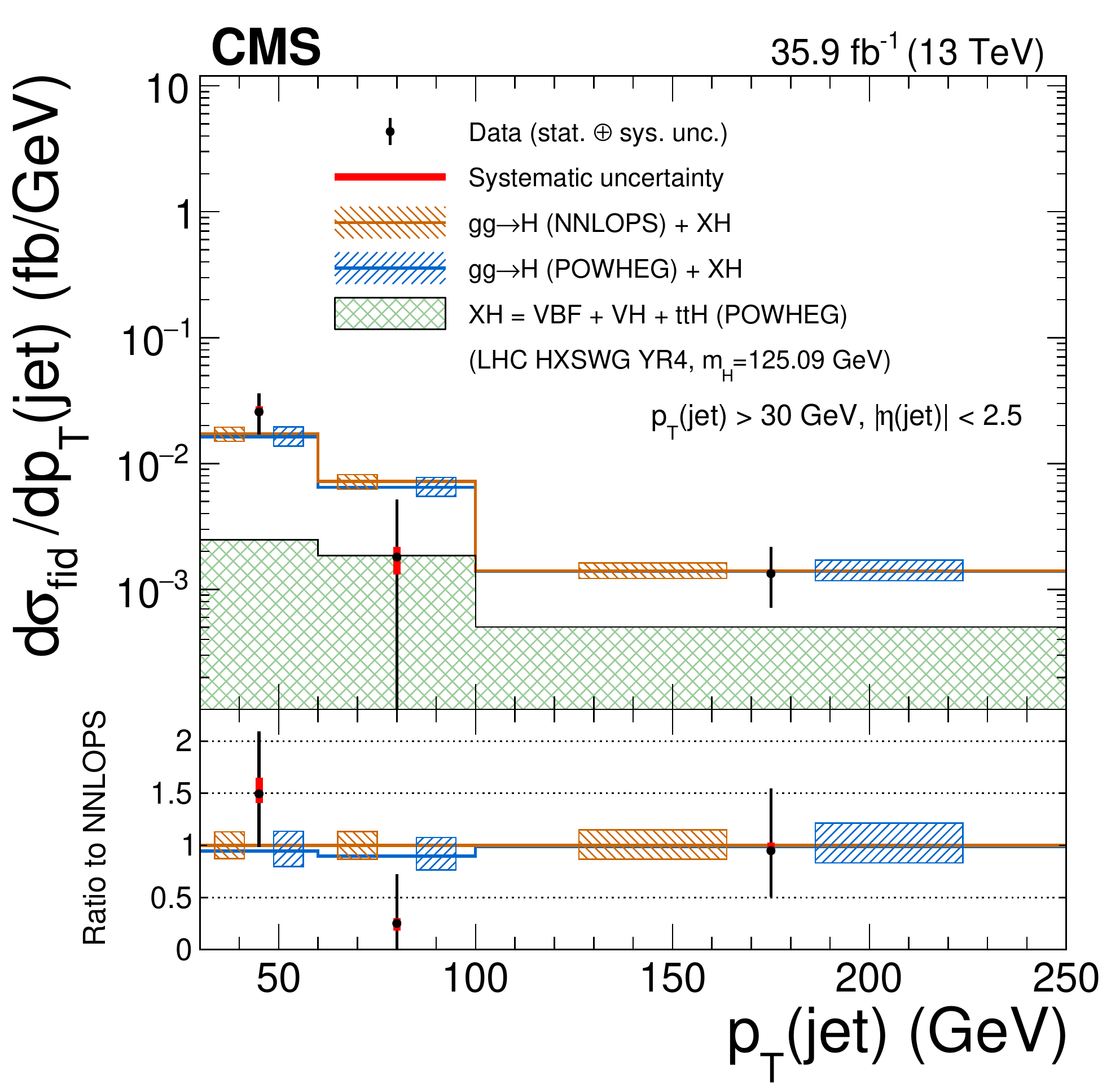}
\includegraphics[width=4.0cm]{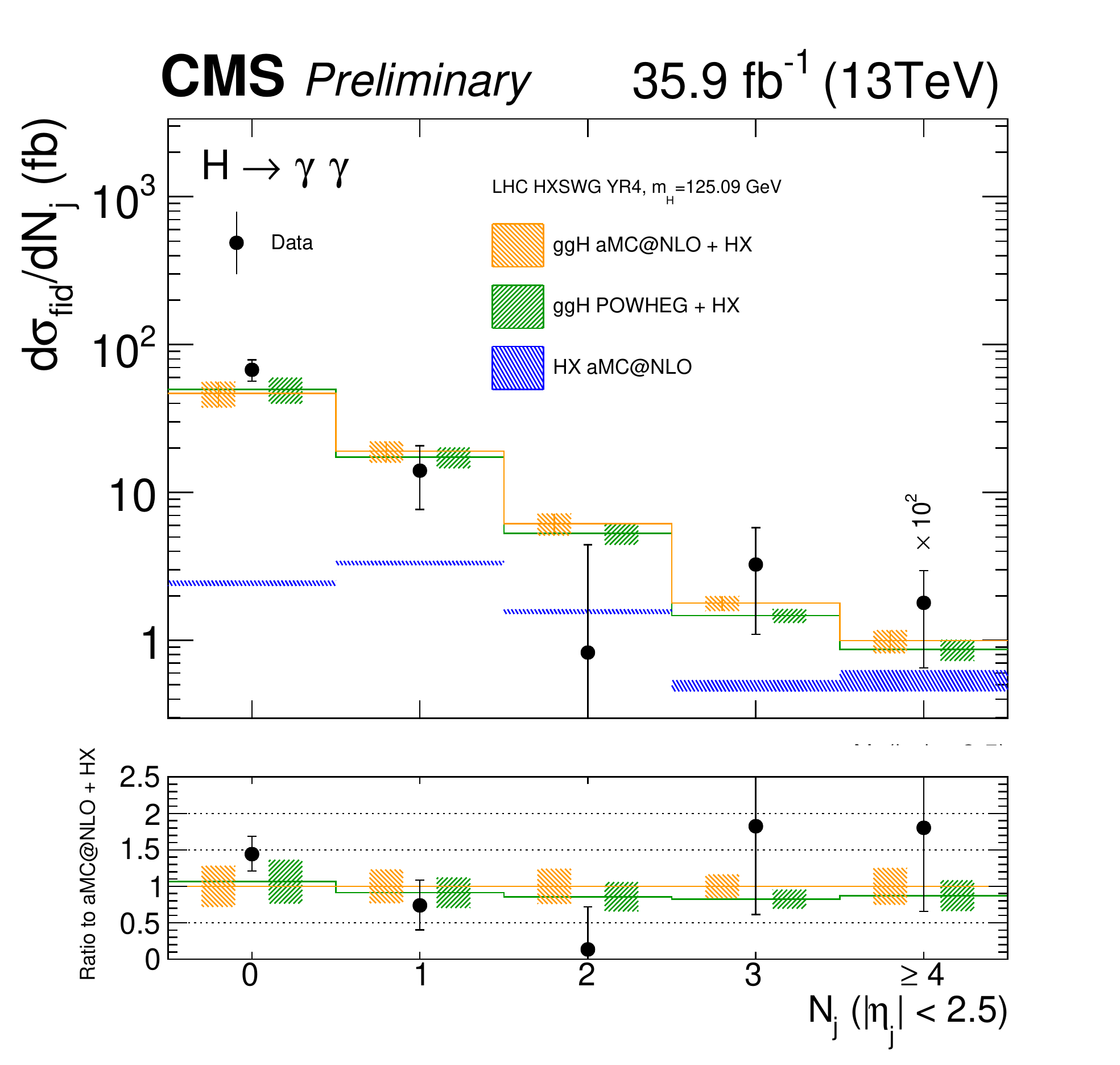}
\caption{ The measured fiducial cross section as a function of $\sqrt{s}$ (top left) with $\mathrm{H}\rightarrow{\rm Z}{\rm Z}\rightarrow4\ell$ decay channel ~\cite{Sirunyan:2017exp}. The results of the differential cross section measurements from $\mathrm{H}\rightarrow{\rm Z}{\rm Z}\rightarrow4\ell$ are shown for $\pt({\rm H})$ (top right) and $\pt({\rm jet})$ of the leading associated jet (bottom left)~\cite{Sirunyan:2017exp}. The differential cross sections as a function of the jet multiplicity N(jets) measured in $\mathrm{H}\rightarrow\gamma\gamma$ channel (bottom right)~\cite{CMS:2017nyv}.
The measurements are compared to the theoretical predictions, combining the Higgs boson
cross sections and branching fraction as in the LHC Higgs Cross Section Working Group~\cite{deFlorian:2016spz} with two different generators for the gluon-gluon fusion process:  NNLOPS~\cite{Hamilton:2013fea} (in orange) and POWHEG~\cite{Alioli:2008gx,Nason:2004rx,Frixione:2007vw} (in green) for $\mathrm{H}\rightarrow{\rm Z}{\rm Z}\rightarrow4\ell$ plots, MADGRAPH\_aMC@NLO~\cite{Alwall:2014hca} (in orange) and POWHEG (in green) for $\mathrm{H}\rightarrow\gamma\gamma$ plot. The subdominant component of the signal (VBF $+$ VH $+~\ttH$) is denoted as XH, and generated using MADGRAPH\_aMC@NLO only and is shown in
blue in the plots.
} \label{fig:HiggsFiducialXS}
\end{figure}

The cross sections for the production and decay of Higgs boson in a tight fiducial phase
space are also measured in both $\mathrm{H}\rightarrow\gamma\gamma$ and $\HZZfl$ decays.
The measured integrated fiducial cross section is compatible to the SM expectation within uncertainties, for each decay channel.
The integrated fiducial cross section measured from $\mathrm{H}\rightarrow{\rm Z}{\rm Z}\rightarrow4\ell$ as a
function of $\sqrt{s}$ is shown in the top left plot of Figure~\ref{fig:HiggsFiducialXS}, compared to the SM
predictions. The differential cross sections as a function of the Higgs boson transverse momentum and the jet multiplicity from both decay channels, and the transverse momentum of the
leading associated jet from $\mathrm{H}\rightarrow{\rm Z}{\rm Z}\rightarrow4\ell$, are also measured and compared with the corresponding theoretical predictions, as shown in the rest plots of Figure~\ref{fig:HiggsFiducialXS}.

\subsection{Higgs-fermion coupling from $\mathrm{H}\rightarrow\tau^{+}\tau^{-}$}

\begin{figure}[htbp]
\centering
\includegraphics[width=3.8cm]{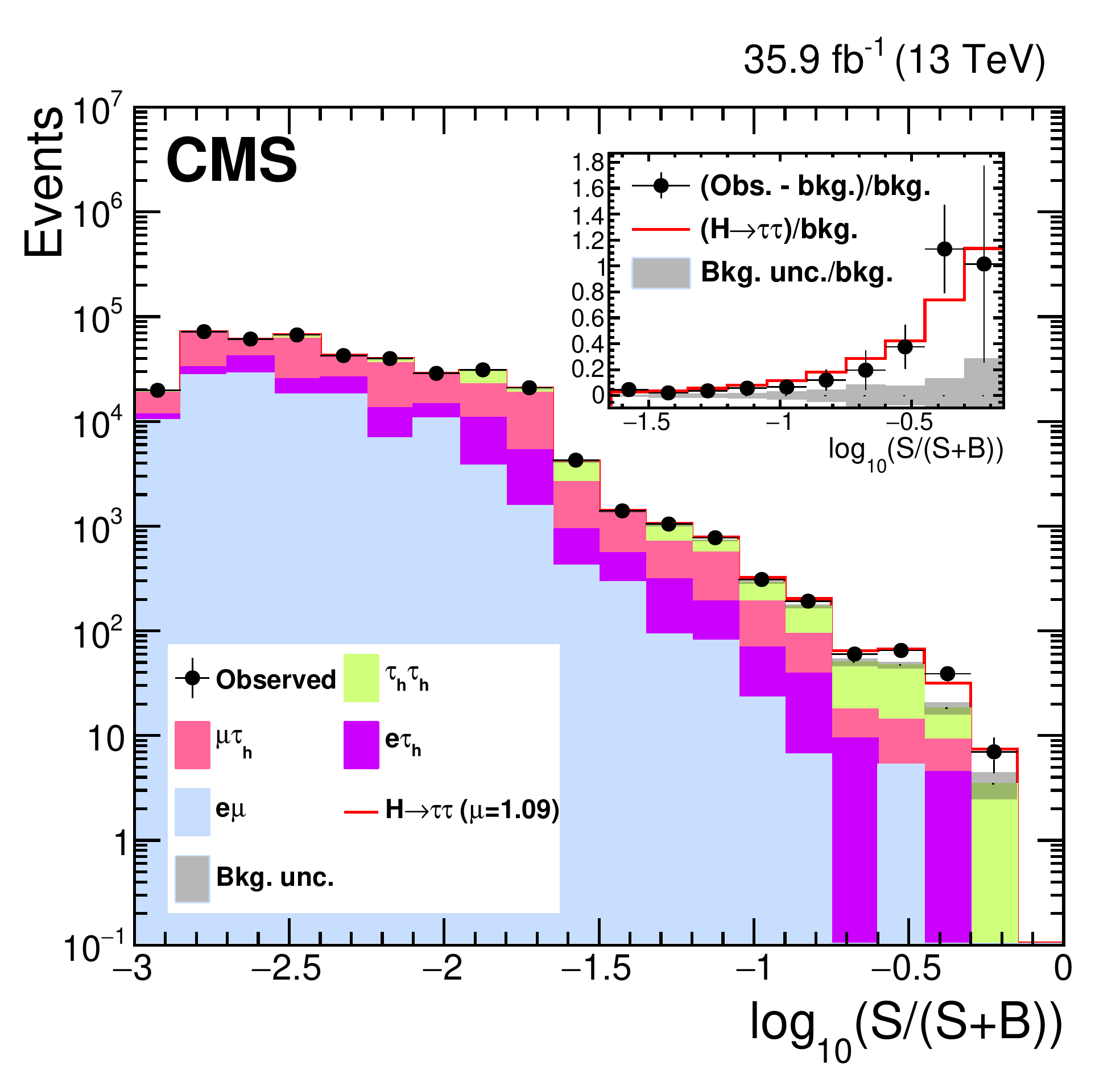}
\includegraphics[width=3.8cm]{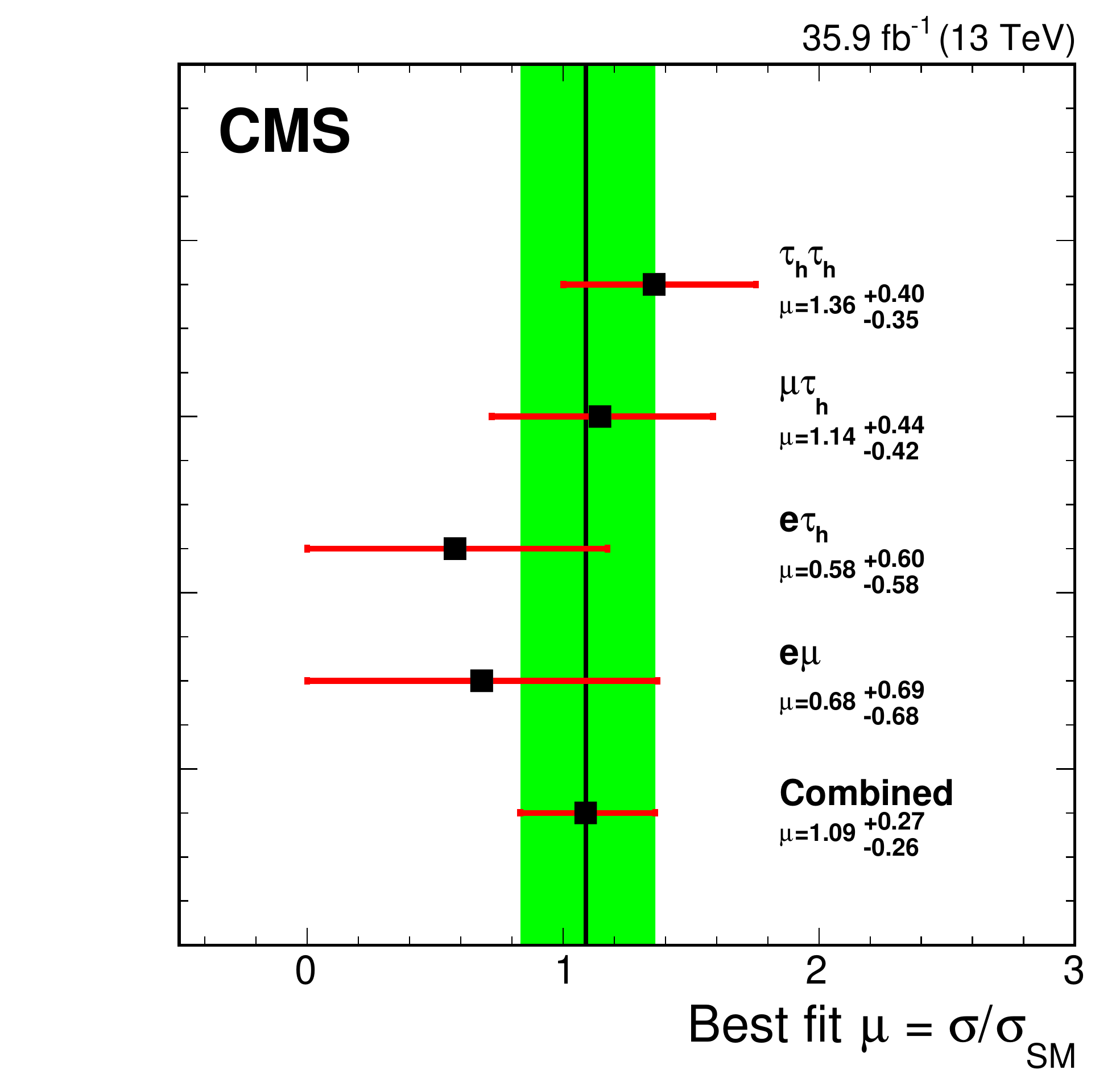}
\caption{ Distribution of the decimal logarithm of the ratio between the expected signal and the sum of expected signal and expected background in each bin of the mass distributions used to extract the results, in all signal regions (left plot)~\cite{Sirunyan:2017khh}. Best fit signal strength per category channel for $\mH = 125.09~\GeV$ (right plot)~\cite{Sirunyan:2017khh}. The constraints from the global fit are used to extract each of the individual best fit signal strengths.
} \label{fig:HiggsTauTau}
\end{figure}

Grouping events in the signal region by their decimal logarithm of the ratio of the signal ($S$) to signal-plus-background ($S+B$) in each bin, an excess of observed events with respect to the SM background expectation is clearly visible in the most sensitive bins of the analysis, as shown in the left plot of Figure~\ref{fig:HiggsTauTau}. The excess in data is quantified by calculating the corresponding local $p$-value using a profile likelihood ratio test statistic. The observed significance for a SM Higgs boson with $\mH = 125.09~\GeV$ is 4.9 standard deviations, for an expected significance of 4.7 standard deviations. The corresponding best fit value for the signal strength $\mu$ is $1.09 ^{+0.27} _{-0.26}$ at $\mH = 125.09~\GeV$, as shown in the right plot of Figure~\ref{fig:HiggsTauTau}. The individual best fit signal strengths per channel, using the constraints obtained on the systematic uncertainties through the global fit, are also given in the right plot of Figure~\ref{fig:HiggsTauTau}.
If combined with the data collected at center-of-mass energies of 7 and 8 TeV~\cite{Khachatryan:2014jba}, both the observed and expected significance are 5.9 standard deviations. The corresponding best fit value for the signal strength $\mu$ is $0.98\pm 0.18$ at $\mH = 125.09~\GeV$. This constitutes the most significant direct measurement of the coupling of the Higgs boson to fermions by a single experiment.


From the search results of Higgs boson decaying to a pair of $\Pgt$ leptons and produced in association with a vector boson~\cite{Sirunyan:2018cpi}, the signal strength is measured relative to the expectation for the standard model Higgs boson, yielding $\mu = 2.5 ^{+1.4} _{-1.3}$.
The results are combined with earlier CMS measurements targeting Higgs boson decays to a pair of $\Pgt$ leptons, performed with the same data set in
the gluon fusion and vector boson fusion production modes.
The combined signal strength is $\mu = 1.24 ^{+0.29} _{-0.27}$ ($1.00 ^{+0.24} _{-0.23}$ expected), and the
observed significance is 5.5 standard deviations (4.8 expected) for a Higgs boson mass of 125~$\GeV$.

\subsection{Observations of $\ttH$ and $\mathrm{H}\rightarrow b\bar{b}$}

\begin{figure}[htbp]
\centering
\includegraphics[width=3.8cm]{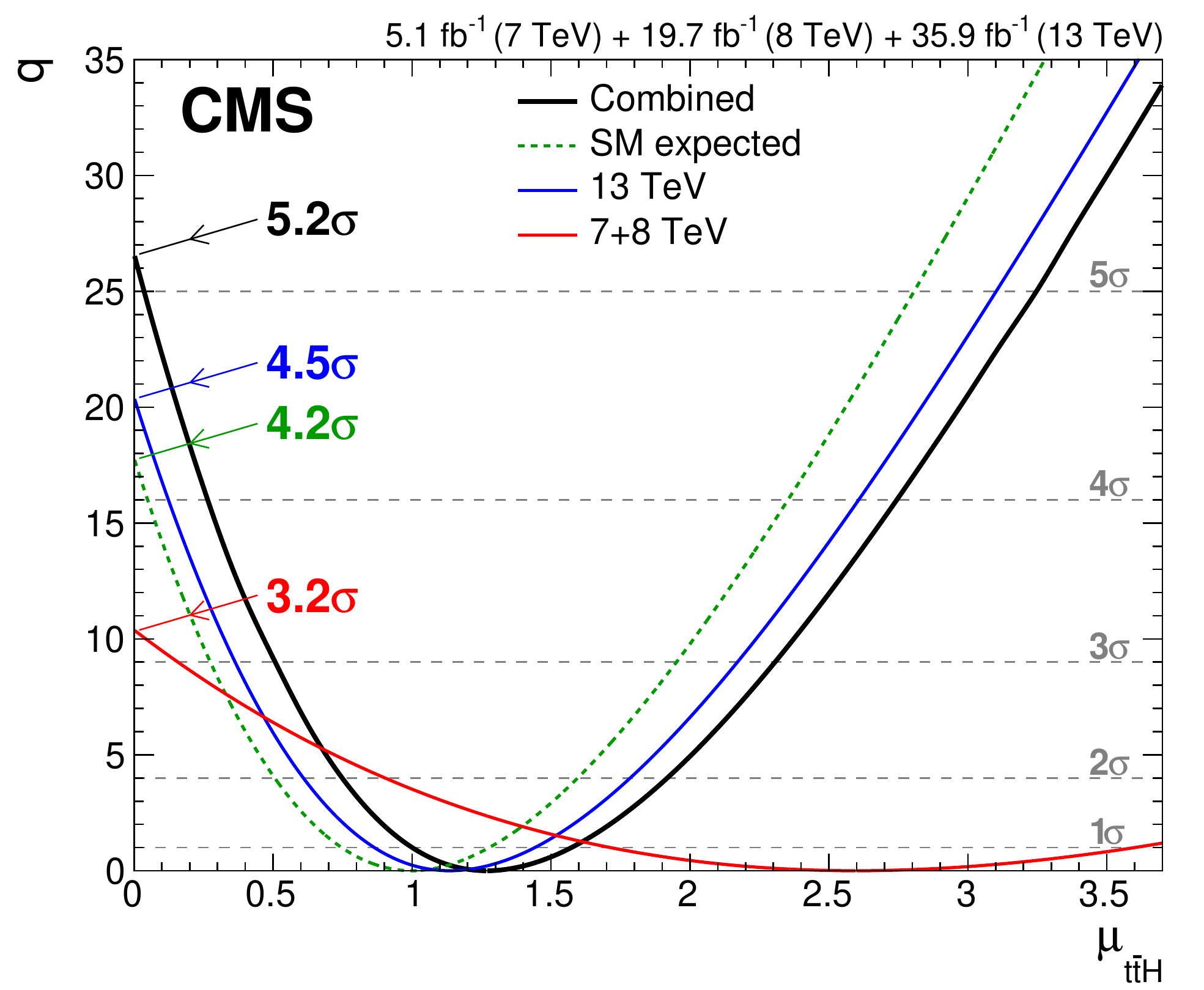}
\includegraphics[width=3.8cm]{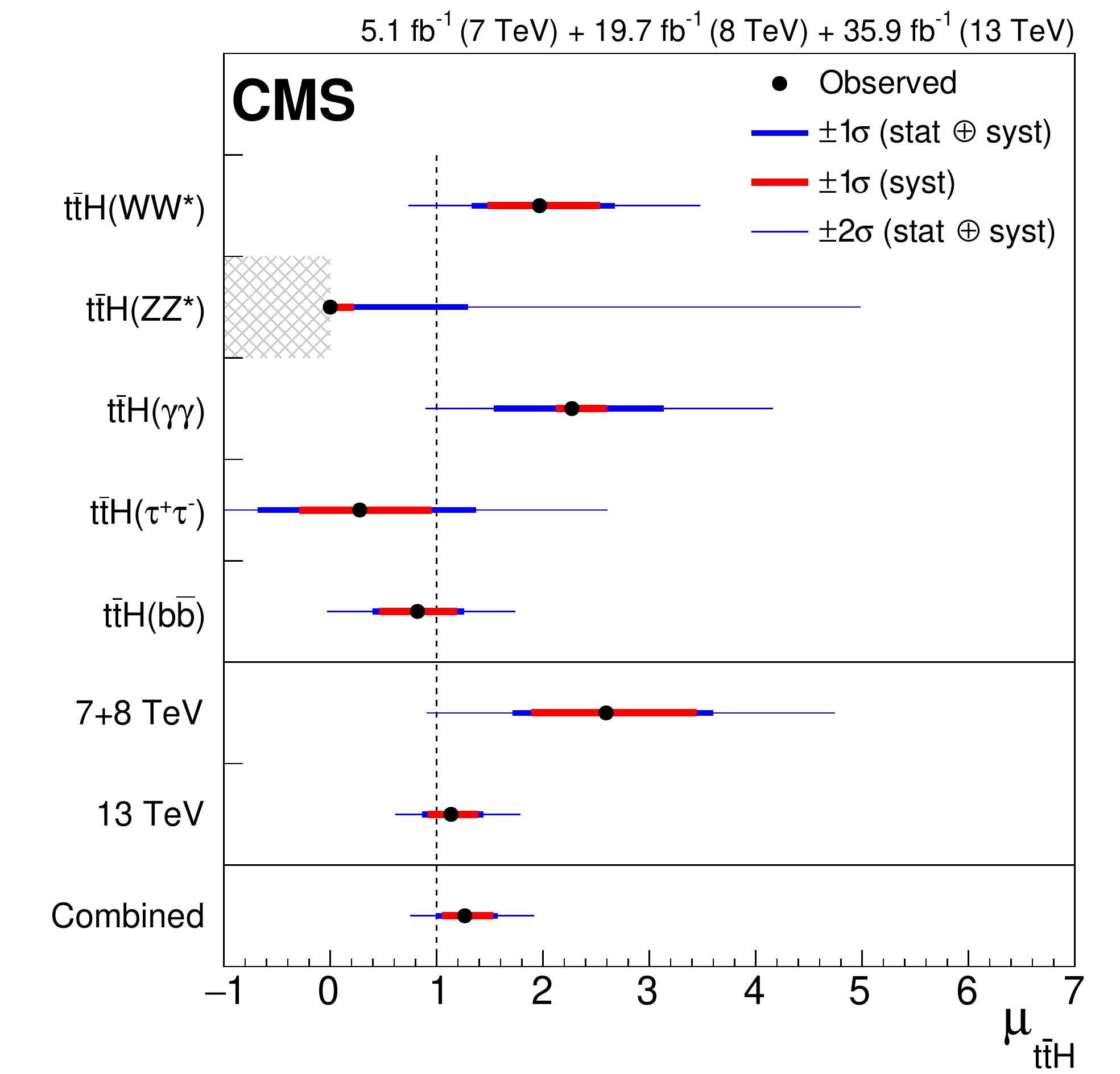}
\caption{ The test statistic $q$, defined as the negative of twice the logarithm of the profile likelihood ratio~\cite{ATLAS:2011tau},
as a function of $\sigmod$
for all decay modes at 7+8$\TeV$ and
at 13$\TeV$, separately,
and for all decay modes at all center-of-mass energies (left plot)~\cite{Sirunyan:2018hoz}.
The expected SM result for the overall combination is also shown.
Best fit value of the $t\bar{t}\PH$ signal strength modifier $\sigmod$ (right plot)~\cite{Sirunyan:2018hoz},
with its 1 and 2 standard deviation confidence intervals ($\sigma$),
for (upper section) the five individual decay channels considered,
(middle section) the combined result for 7+8$\TeV$ alone
and for 13$\TeV$ alone,
and (lower section) the overall combined result.
The Higgs boson mass is taken to be 125.09~$\GeV$.
} \label{fig:ttHObs}
\end{figure}

Higgs boson production in association with a top quark-antiquark pair is observed, based on a combined analysis of proton-proton collision data at
center-of-mass energies of $\sqrt{s}=7$, 8, and 13~$\TeV$,
corresponding to integrated luminosities of up to
5.1, 19.7, and 35.9~$\fbinv$, respectively~\cite{Sirunyan:2018hoz}.
The results of statistically independent searches for Higgs bosons
produced in conjunction with a top quark-antiquark pair
and decaying to pairs of $\PW$ bosons, $\cPZ$ bosons, photons, $\Pgt$ leptons,
or bottom quark jets are combined to maximize sensitivity.
An excess of events is observed,
with a significance of 5.2 standard deviations,
over the expectation from the background-only hypothesis.
The corresponding expected significance from the standard model
for a Higgs boson mass of 125.09~$\GeV$ is 4.2 standard deviations.
The combined best fit signal strength normalized to the standard model prediction
is $1.26{^{+0.31}_{-0.26}}$. The results are shown in Figure~\ref{fig:ttHObs}.

The search for Higgs boson  produced in association with an electroweak
vector boson and decaying to a $\bbbar$ pair is performed. When combined all $\VH$ measurements using data collected at $\sqrt{s}=7$, 8, and 13~$\TeV$, an excess of events is observed at $\mH = 125.09~\GeV$ with a significance of 4.8 standard deviations, where the expectation for the SM Higgs boson is 4.9.  The corresponding measured signal strength is $1.01\pm 0.22$. The results are summarized in Table~\ref{tab:Hbb_limits_by_mode}. Combining this result with previous measurements by the CMS Collaboration of the $\HBB$ decay in events where the Higgs boson is produced through gluon fusion, vector boson fusion, or in association with top quarks, the observed\,(expected) significance increases to 5.6\,(5.5) standard deviations and the signal strength is $\mu = 1.04\pm 0.20$. This constitutes the observation of the $\HBB$ decay by the CMS Collaboration.

\begin{center}
\begin{table}[hbt]
\setlength{\tabcolsep}{0.2pc}
\begin{tabular}{lccc}
\hline
                  & \multicolumn{2}{c}{Significance ($\sigma$)}   &                        \\
Data set          & Expected      & Observed           & Signal strength        \\
\hline
2017              & $3.1$         &  $3.3$             &  $1.08\pm 0.34$                      \\
 Run 2            & $4.2$         &  $4.4$             & $1.06\pm 0.26$ \\
 Run 1 + Run 2    & $4.9$         &  $4.8$             & $1.01\pm 0.22$ \\
\hline
\end{tabular}
\caption{Expected and observed significances, in $\sigma$, and observed signal strengths for the $\VH$ production process with $\HBB$~\cite{Sirunyan:2018kst}. Results are shown separately for 2017 data, combined Run 2 (2016 and 2017) data, and for the combination of the Run 1 and Run 2 data sets.
All results are obtained for $\mH=125.09~\GeV$ combining statistical and systematic uncertainties.}
\label{tab:Hbb_limits_by_mode}
\end{table}
\end{center}

\subsection{Upper limits on the Higgs boson rare decays}

\begin{figure}[htbp]
\centering
\includegraphics[width=3.8cm]{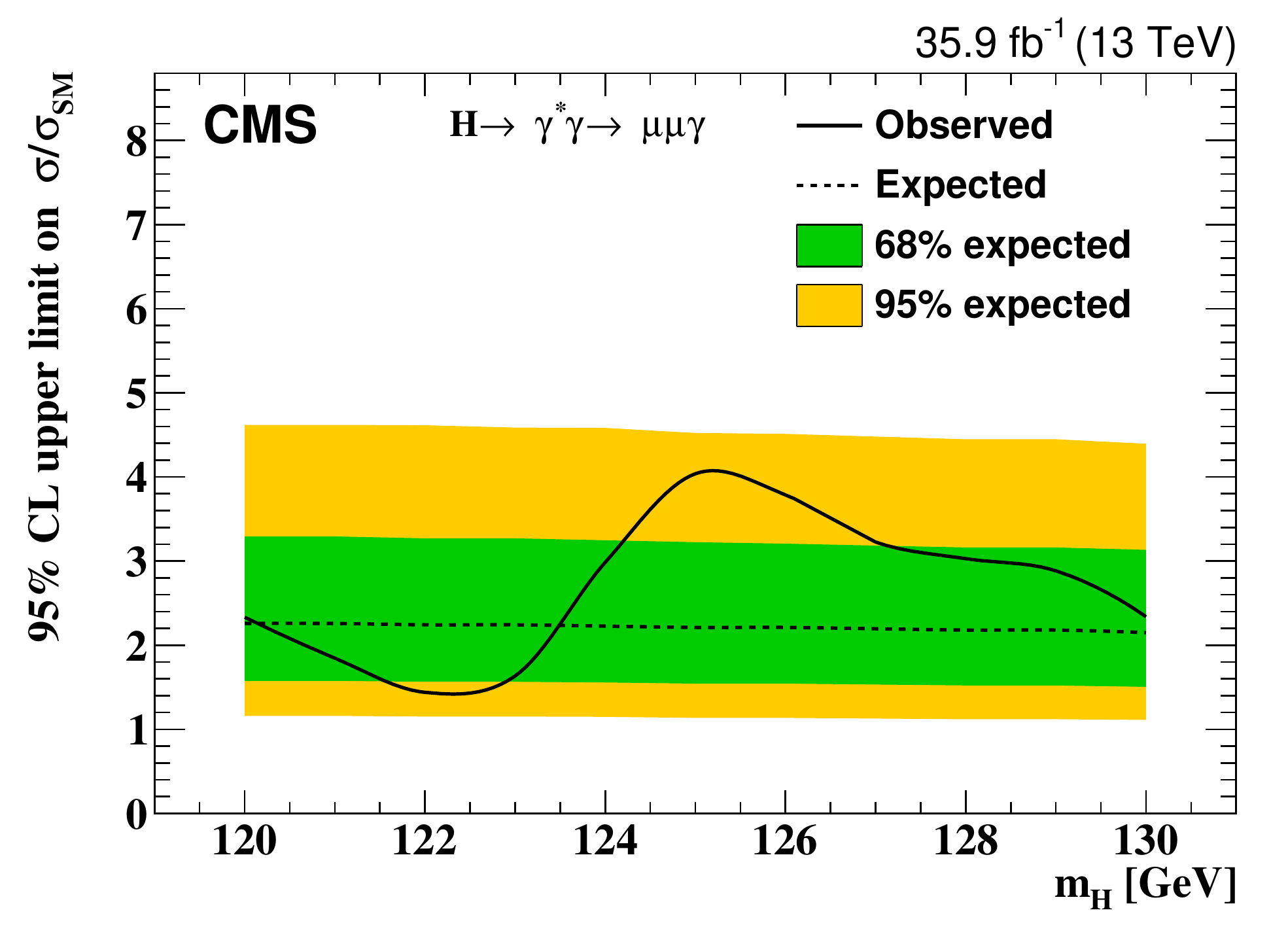}
\includegraphics[width=3.8cm]{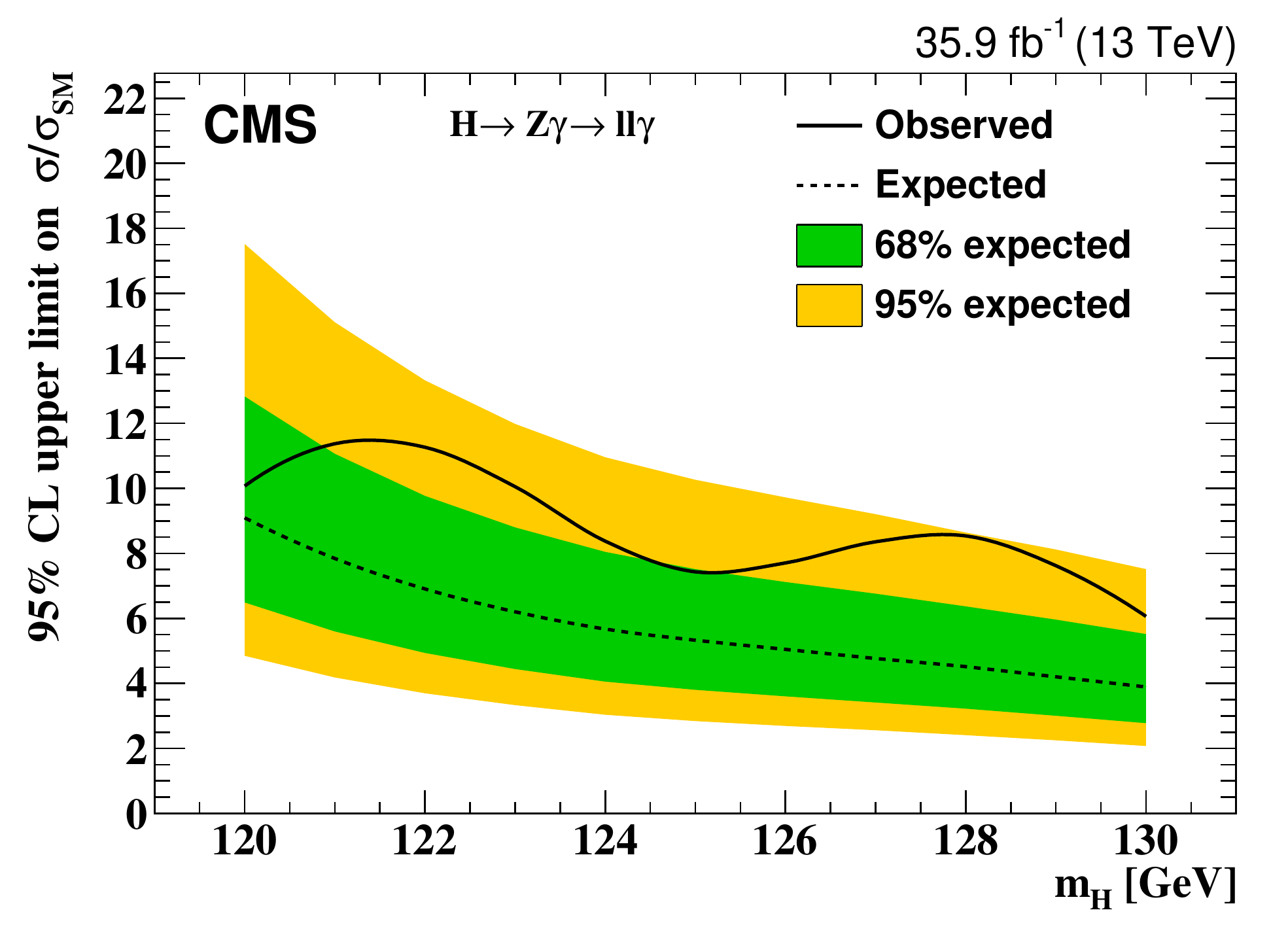}
\caption{ Exclusion limit, at 95\% $\CL$, on the cross section of the $\PH\to\gamma^*\gamma\to\mu\mu\gamma$
 process (left plot) and the $\PH\to\cPZ\gamma\to\ell\ell\gamma$ process (right plot)
relative to the SM prediction, as a function of the Higgs boson mass~\cite{Sirunyan:2018tbk}.
} \label{fig:Hllgamma}
\end{figure}

For Higgs boson decays to a $\cPZ$ boson and a photon
($\PH\to\cPZ\gamma\to\ell\ell\gamma$,$\ell=\Pe$ or $\mu$),
or to two photons, one of which has an internal conversion
into a muon pair ($\PH\to\gamma^{*}\gamma\to\mu\mu\gamma$), no significant excess above the
expected background is found from 2016 data samples. Limits on the Higgs boson
production cross section times the corresponding branching fractions
are set. The expected exclusion limits at 95\% confidence level
are about 2.1--2.3 (3.9--9.1) times the SM cross section in the
$\PH\to\gamma^*\gamma\to\mu\mu\gamma$
($\PH\to\cPZ\gamma\to\ell\ell\gamma$) channel in the mass range from
120 to 130~$\GeV$, and the observed limit varies between about 1.4 and 4.0
(6.1 and 11.4) times the SM cross section. Finally, the
$\PH\to\gamma^*\gamma\to\mu\mu\gamma$ and
$\PH\to\cPZ\gamma\to\ell\ell\gamma$ analyses are combined for
$m_\PH=125~\GeV$, obtaining an observed (expected) 95$\%$ confidence level upper limit
of 3.9 (2.0) times the SM cross section. The results are shown in Figure~\ref{fig:Hllgamma}.

For the Higgs boson decaying to two muons~\cite{Sirunyan:2018hbu},
the 95\% confidence level observed (expected) upper limit
on the production cross section times branching fraction to a pair of muons is found
to be 2.95 (2.45) times the standard model expectation at the mass of $125.09~\GeV$, using 2016 data.
In combination with Run~1 data,
the observed (expected) upper
limit improves to 2.92 (2.16) times the standard model value.
This corresponds to
an upper limit on the standard model Higgs boson branching fraction to muons of
$6.4\times10^{-4}$ .

\subsection{Combinations on signal strength and couplings}

\begin{figure}[htbp]
\centering
\includegraphics[width=3.5cm]{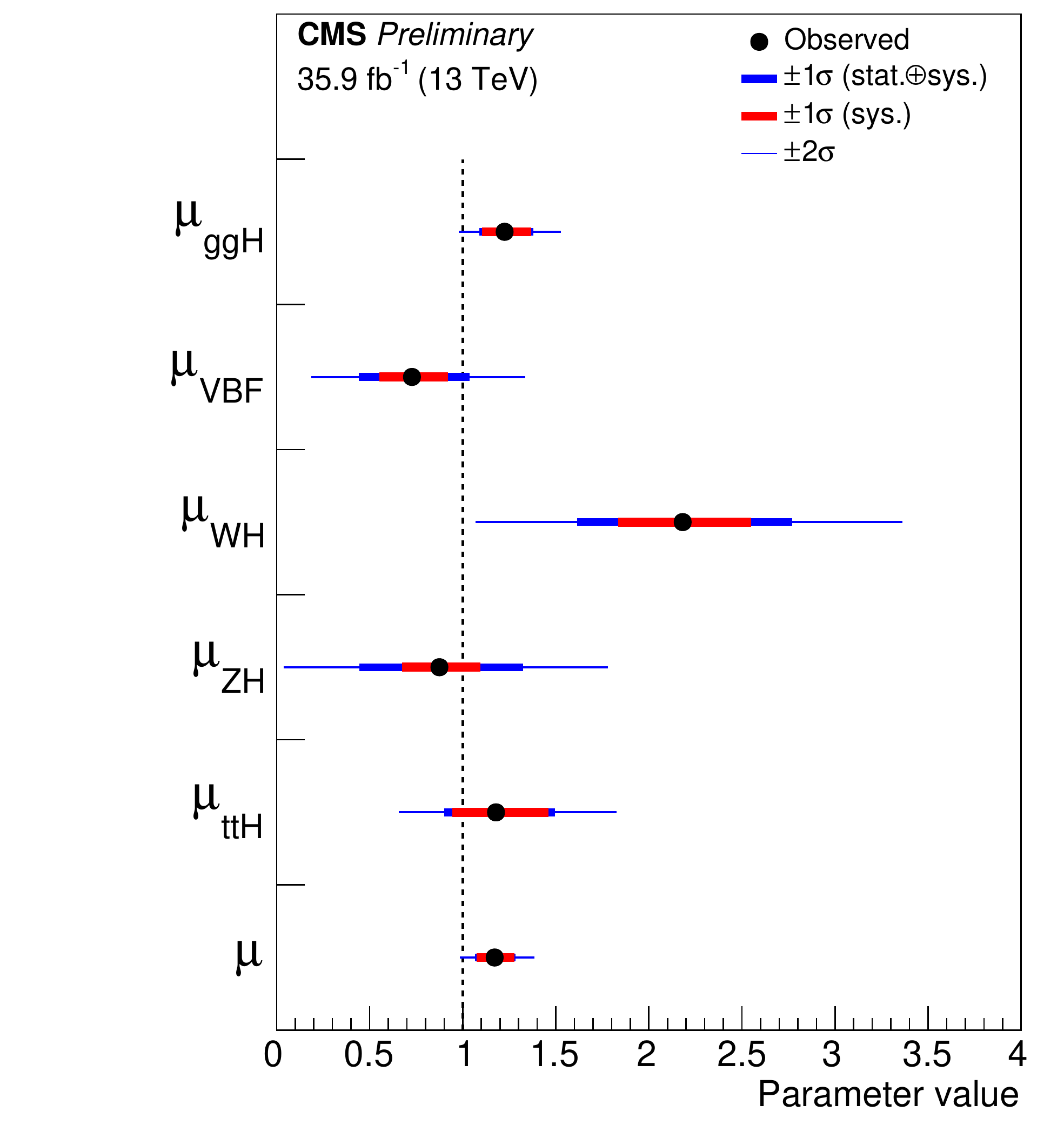}
\includegraphics[width=3.8cm]{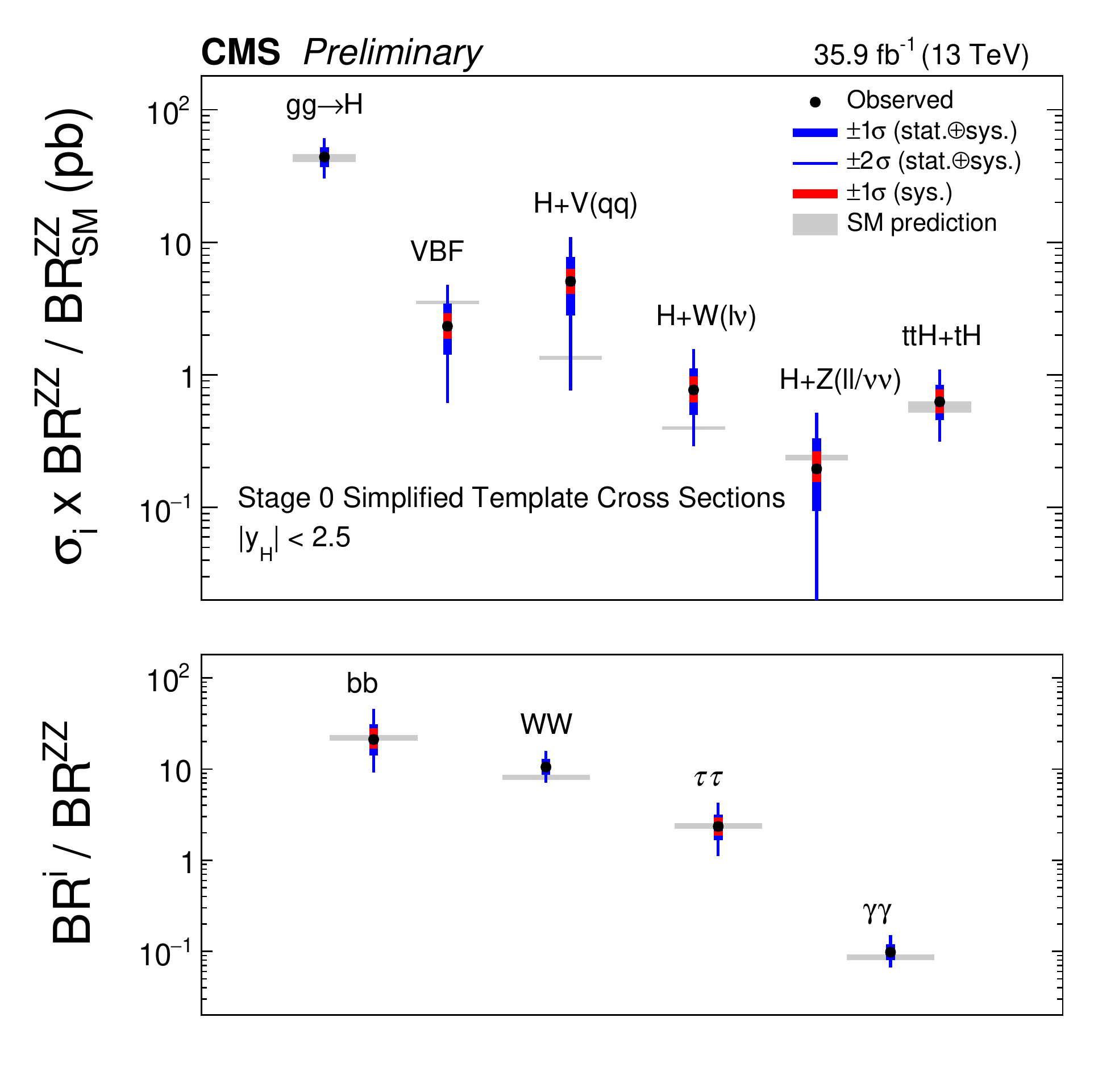}
\includegraphics[width=3.8cm]{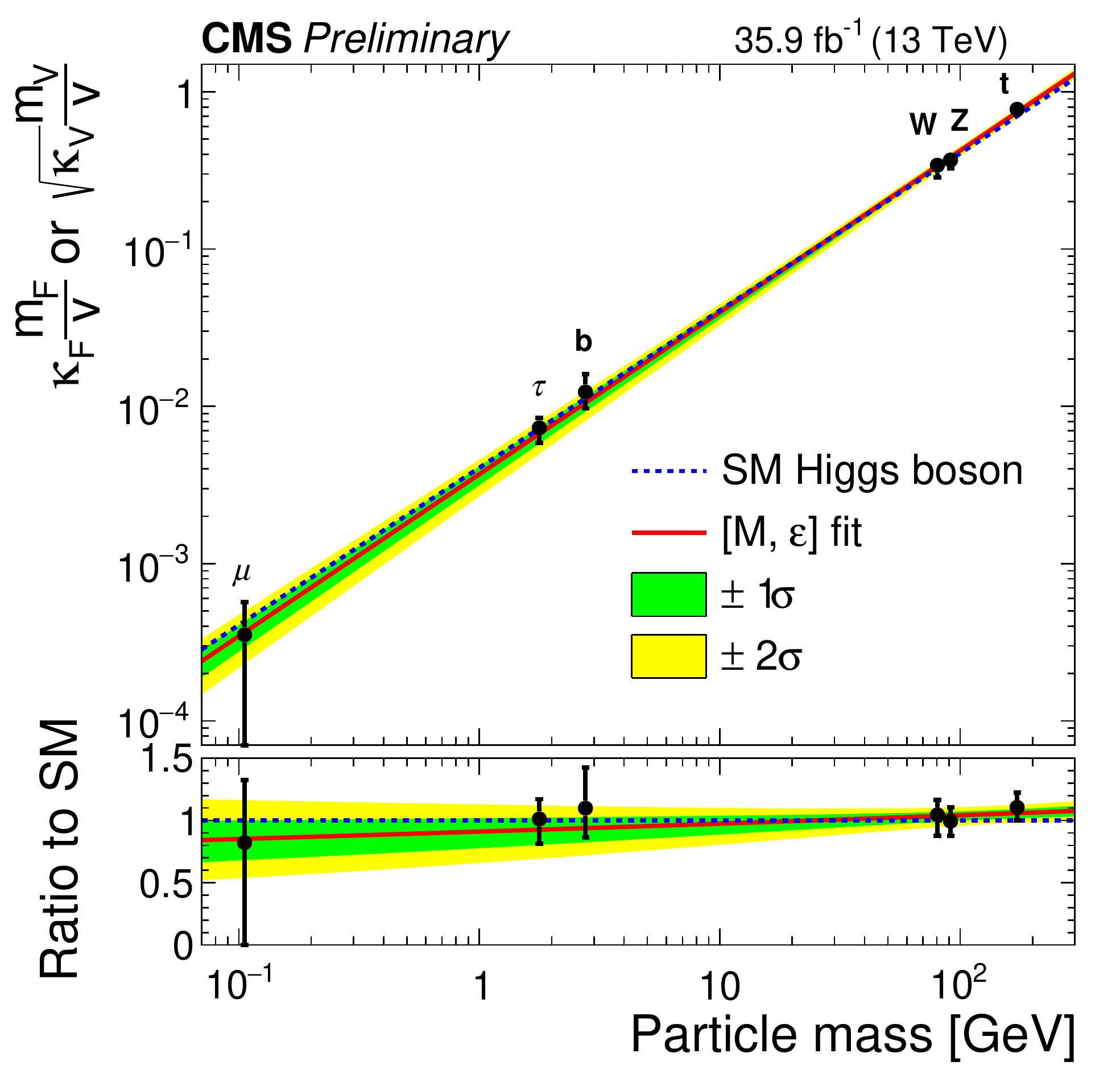}
\includegraphics[width=3.8cm]{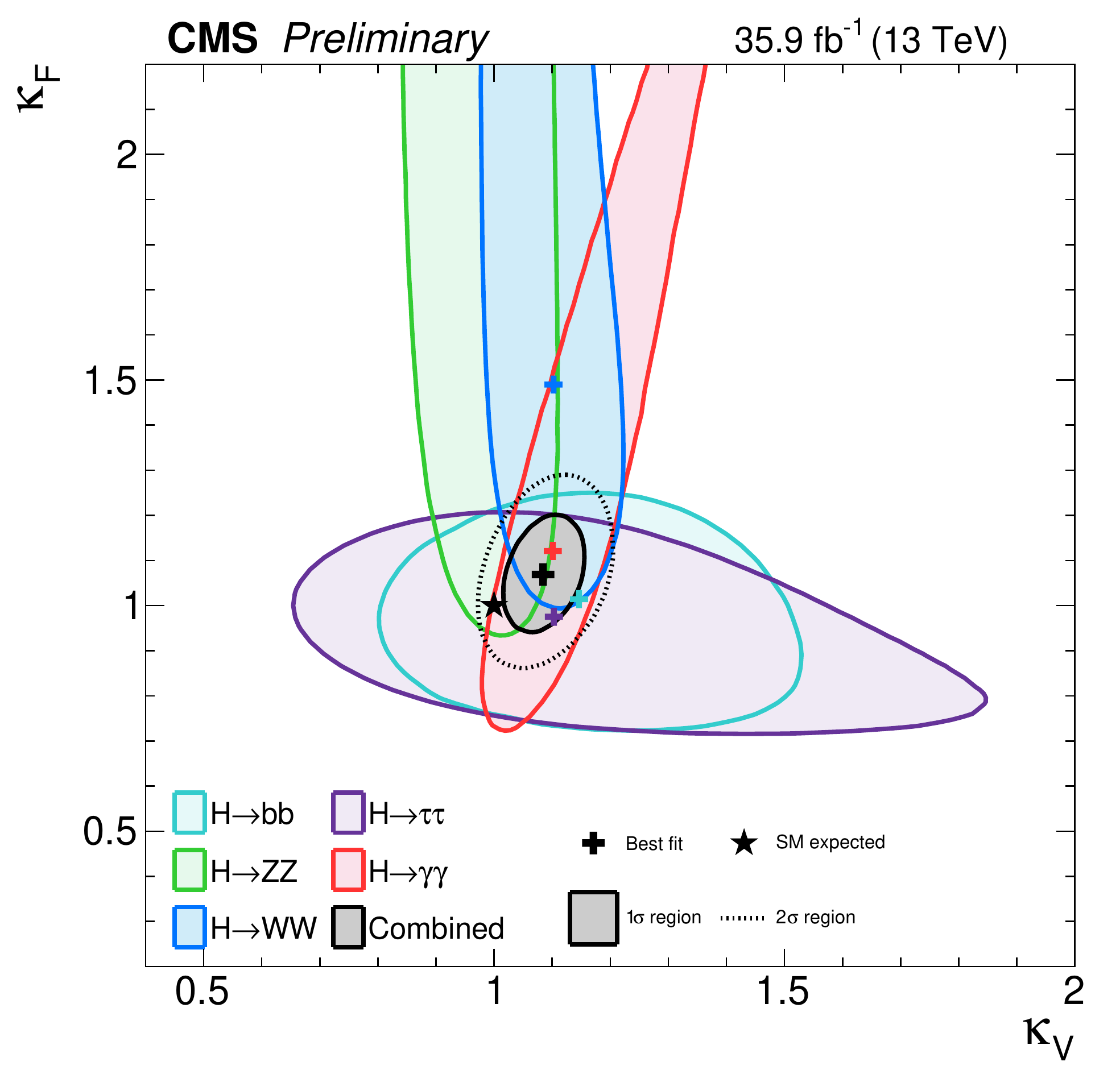}
\caption{ Summary plots of the fit to the per-production mode signal strength modifiers (top left), the stage 0 simplified template cross sections as the ratios of cross sections and branching fractions (top right), results of the fit using the six parameter $\kappa$-framework model plotted versus the particle masses overlayed with the phenomenological $(M,\epsilon)$ fit ($\kappa_{F} = v \; m_\mathrm{f}^{\epsilon} / M^{1+\epsilon}$
for fermions and $\kappa_{V} = v \; m_\mathrm{V}^{2\epsilon} / M^{1+2\epsilon}$
for vector bosons, with the SM Higgs boson vacuum expectation value $v=246.22$~$\GeV$) for comparison (bottom left),
and the $1\sigma$ and $2\sigma$ CL regions in the $\kappa_{\rm{F}}$ vs.
$\kappa_{\rm{V}}$ parameter space for the model assuming a common scaling of all the vector boson and fermion couplings (bottom right)~\cite{CMS:2018lkl}.
} \label{fig:HiggsCombined}
\end{figure}

Selected results of the combined measurements of the production and decay rates of the Higgs boson~\cite{CMS:2018lkl}, as well as its couplings to vector bosons and fermions, are shown in Figure~\ref{fig:HiggsCombined}.
The best-fit ratio of the signal yield to the standard model expectation is measured to be $\mu=1.17\pm0.10 = 1.17\pm{0.06}~\mathrm{(stat.)}~^{+0.06}_{-0.05}~\mathrm{(sig.~theo.)}~\pm{0.06}~\mathrm{(other~syst.)}$, assuming a Higgs boson mass of $125.09\,~\mathrm{GeV}$.
An improvement in the measured precision of the gluon fusion production rate of around $\sim$50\% is achieved compared to previous ATLAS and CMS measurements~\cite{Khachatryan:2016vau}.
Additional results are given for parametrizations with varying assumptions on the scaling behavior of the different production and decay modes, including generic ones based on ratios of cross sections and branching fractions or coupling modifiers. The results are compatible with the standard model predictions in all parametrizations considered.

\section{Conclusions}

The results of the measurements of the 125~$\GeV$ Higgs boson properties at CMS are presented.
The measured Higgs boson properties include its mass, signal strength relative to the standard model prediction, signal strength modifiers for different Higgs boson production modes, coupling modifiers to fermions and bosons, effective coupling modifiers to photons and gluons, simplified template cross sections, total and differential fiducial cross sections. All results are consistent, within their uncertainties, with the expectations for the Standard Model Higgs boson. Many achievements beyond the Higgs boson discovery have been obtained at the CMS, mainly based on the pp collision data
at $\sqrt{s}=13~\TeV$. The achievements include the improved precision in Higgs boson properties, observation of $\ttH$  production and single-experiment observation of $\mathrm{H}\rightarrow\tau^{+}\tau^{-}$ by CMS, observation of $\mathrm{H}\rightarrow b\bar{b}$ decay.

\Acknowledgements

I would like to thank the QCD2018 organizers for their hospitality and
the wonderful working environment. I acknowledge the support from
National Natural Science Foundation of China (No. 11505208, No. 11875275 and No. 11661141007) and
China Ministry of Science and Technology (No. 2018YFA0403901).


\begin{thebibliography}{99}
\vspace*{-0.25cm}

\bibitem{Glashow:1961tr}
  S.~L.~Glashow,
  Nucl.\ Phys.\  {\bf 22} (1961) 579.
  doi:10.1016/0029-5582(61)90469-2

\bibitem{Weinberg:1967tq}
  S.~Weinberg,
  Phys.\ Rev.\ Lett.\  {\bf 19} (1967) 1264.
  doi:10.1103/PhysRevLett.19.1264

\bibitem{Salam:1968rm}
  A.~Salam,
  Conf.\ Proc.\ C {\bf 680519} (1968) 367.

\bibitem{Aad:2012tfa}
  G.~Aad {\it et al.} [ATLAS Collaboration],
  Phys.\ Lett.\ B {\bf 716} (2012) 1
  doi:10.1016/j.physletb.2012.08.020
  [arXiv:1207.7214 [hep-ex]].

\bibitem{Chatrchyan:2012xdj}
  S.~Chatrchyan {\it et al.} [CMS Collaboration],
  Phys.\ Lett.\ B {\bf 716} (2012) 30
  doi:10.1016/j.physletb.2012.08.021
  [arXiv:1207.7235 [hep-ex]].

\bibitem{Chatrchyan:2013lba}
  S.~Chatrchyan {\it et al.} [CMS Collaboration],
  JHEP {\bf 1306} (2013) 081
  doi:10.1007/JHEP06(2013)081
  [arXiv:1303.4571 [hep-ex]].

\bibitem{Aad:2015zhl}
  G.~Aad {\it et al.} [ATLAS and CMS Collaborations],
  Phys.\ Rev.\ Lett.\  {\bf 114} (2015) 191803
  doi:10.1103/PhysRevLett.114.191803
  [arXiv:1503.07589 [hep-ex]].

\bibitem{Khachatryan:2016vau}
  G.~Aad {\it et al.} [ATLAS and CMS Collaborations],
  JHEP {\bf 1608} (2016) 045
  doi:10.1007/JHEP08(2016)045
  [arXiv:1606.02266 [hep-ex]].

\bibitem{Khachatryan:2014kca}
  V.~Khachatryan {\it et al.} [CMS Collaboration],
  Phys.\ Rev.\ D {\bf 92} (2015) no.1,  012004
  doi:10.1103/PhysRevD.92.012004
  [arXiv:1411.3441 [hep-ex]].

\bibitem{Khachatryan:2016ctc}
  V.~Khachatryan {\it et al.} [CMS Collaboration],
  JHEP {\bf 1609} (2016) 051
  doi:10.1007/JHEP09(2016)051
  [arXiv:1605.02329 [hep-ex]].


\bibitem{Englert:1964et}
  F.~Englert and R.~Brout,
  Phys.\ Rev.\ Lett.\  {\bf 13} (1964) 321.
  doi:10.1103/PhysRevLett.13.321


\bibitem{Higgs:1964pj}
  P.~W.~Higgs,
  Phys.\ Rev.\ Lett.\  {\bf 13} (1964) 508.
  doi:10.1103/PhysRevLett.13.508


\bibitem{Guralnik:1964eu}
  G.~S.~Guralnik, C.~R.~Hagen and T.~W.~B.~Kibble,
  Phys.\ Rev.\ Lett.\  {\bf 13} (1964) 585.
  doi:10.1103/PhysRevLett.13.585

\bibitem{Chatrchyan:2008aa}
  S.~Chatrchyan {\it et al.} [CMS Collaboration],
  JINST {\bf 3} (2008) S08004.
  doi:10.1088/1748-0221/3/08/S08004

\bibitem{CMS:2017rli}
  CMS Collaboration, 
  CMS-PAS-HIG-16-040,
  http://cds.cern.ch/record/2264515.

\bibitem{CMS:2017nyv}
  CMS Collaboration, 
  CMS-PAS-HIG-17-015,
  http://cds.cern.ch/record/2257530.

\bibitem{Sirunyan:2017exp}
  A.~M.~Sirunyan {\it et al.} [CMS Collaboration],
  JHEP {\bf 1711} (2017) 047
  doi:10.1007/JHEP11(2017)047
  [arXiv:1706.09936 [hep-ex]].

\bibitem{Sirunyan:2017tqd}
  A.~M.~Sirunyan {\it et al.} [CMS Collaboration],
  Phys.\ Lett.\ B {\bf 775} \\ (2017) 1
  doi:10.1016/j.physletb.2017.10.021
  [arXiv:1707.00541 [hep-ex]].

\bibitem{CMS:2018mmw}
  CMS Collaboration, 
  CMS-PAS-HIG-18-001,
  http://cds.cern.ch/record/2621419.

\bibitem{Sirunyan:2018egh}
  A.~M.~Sirunyan {\it et al.} [CMS Collaboration],
  [arXiv:1806.05246 [hep-ex]].

\bibitem{Sirunyan:2017khh}
  A.~M.~Sirunyan {\it et al.} [CMS Collaboration],
  Phys.\ Lett.\ B {\bf 779} (2018) 283
  doi:10.1016/j.physletb.2018.02.004
  [arXiv:1708.00373 [hep-ex]].

\bibitem{Sirunyan:2018cpi}
  A.~M.~Sirunyan {\it et al.} [CMS Collaboration],
  arXiv:1809.03590 [hep-ex].


\bibitem{Sirunyan:2017elk}
  A.~M.~Sirunyan {\it et al.} [CMS Collaboration],
  Phys.\ Lett.\ B {\bf 780} (2018) 501
  doi:10.1016/j.physletb.2018.02.050
  [arXiv:1709.07497 [hep-ex]].

\bibitem{Sirunyan:2018kst}
  A.~M.~Sirunyan {\it et al.} [CMS Collaboration],
  Phys.\ Rev.\ Lett.\ {\bf 121} (2018) 121801
  doi:10.1103/PhysRevLett.121.121801
  [arXiv:1808.08242 [hep-ex]].



\bibitem{Sirunyan:2018tbk}
  A.~M.~Sirunyan {\it et al.} [CMS Collaboration],
  [arXiv:1806.05996 [hep-ex]].

\bibitem{Sirunyan:2018hbu}
  A.~M.~Sirunyan {\it et al.} [CMS Collaboration],
  arXiv:1807.06325 [hep-ex].

\bibitem{CMS:2018lkl}
  CMS Collaboration, 
  CMS-PAS-HIG-17-031,
  http://cds.cern.ch/record/2308127.

\bibitem{deFlorian:2016spz}
  D.~de Florian {\it et al.} [LHC Higgs Cross Section Working Group],
  doi:10.23731/CYRM-2017-002
  arXiv:1610.07922 [hep-ph].



\bibitem{Dauncey:2014xga}
  P.~D.~Dauncey, M.~Kenzie, N.~Wardle and G.~J.~Davies,
  JINST {\bf 10} (2015) no.04,  P04015
  doi:10.1088/1748-0221/10/04/P04015
  [arXiv:1408.6865 [physics.data-an]].



\bibitem{Khachatryan:2015cwa}
  V.~Khachatryan {\it et al.} [CMS Collaboration],
  JHEP {\bf 1510} (2015) 144
  doi:10.1007/JHEP10(2015)144
  [arXiv:1504.00936 [hep-ex]].

\bibitem{Khachatryan:2015mma}
  V.~Khachatryan {\it et al.} [CMS Collaboration],
  Phys.\ Rev.\ D {\bf 92} (2015) no.7,  072010
  doi:10.1103/PhysRevD.92.072010
  [arXiv:1507.06656 [hep-ex]].








\bibitem{Nambu:1961fr}
  Y.~Nambu and G.~Jona-Lasinio,
  Phys.\ Rev.\  {\bf 124} (1961) 246.
  doi:10.1103/PhysRev.124.246


\bibitem{Chatrchyan:2013zna}
  S.~Chatrchyan {\it et al.} [CMS Collaboration],
  Phys.\ Rev.\ D {\bf 89} (2014) no.1,  012003
  doi:10.1103/PhysRevD.89.012003
  [arXiv:1310.3687 [hep-ex]].

\bibitem{Khachatryan:2015bnx}
  V.~Khachatryan {\it et al.} [CMS Collaboration],
  Phys.\ Rev.\ D {\bf 92} (2015) no.3,  032008
  doi:10.1103/PhysRevD.92.032008
  [arXiv:1506.01010 [hep-ex]].



\bibitem{Chen:2012ju}
  L.~B.~Chen, C.~F.~Qiao and R.~L.~Zhu,
  Phys.\ Lett.\ B {\bf 726} (2013) 306
  doi:10.1016/j.physletb.2013.08.050
  [arXiv:1211.6058 [hep-ph]].

\bibitem{Sun:2013rqa}
  Y.~Sun, H.~R.~Chang and D.~N.~Gao,
  JHEP {\bf 1305} (2013) 061
  doi:10.1007/JHEP05(2013)061
  [arXiv:1303.2230 [hep-ph]].

\bibitem{Passarino:2013nka}
  G.~Passarino,
  Phys.\ Lett.\ B {\bf 727} (2013) 424
  doi:10.1016/j.physletb.2013.10.052
  [arXiv:1308.0422 [hep-ph]].

















\bibitem{Patrignani:2016xqp}
  C.~Patrignani {\it et al.} [Particle Data Group],
  Chin.\ Phys.\ C {\bf 40} (2016) no.10,  100001.
  doi:10.1088/1674-1137/40/10/100001

\bibitem{Hamilton:2013fea}
  K.~Hamilton, P.~Nason, E.~Re and G.~Zanderighi,
  JHEP {\bf 1310} (2013) 222
  doi:10.1007/JHEP10(2013)222
  [arXiv:1309.0017 [hep-ph]].


\bibitem{Alioli:2008gx}
  S.~Alioli, P.~Nason, C.~Oleari and E.~Re,
  JHEP {\bf 0807} (2008) 060
  doi:10.1088/1126-6708/2008/07/060
  [arXiv:0805.4802 [hep-ph]].

\bibitem{Nason:2004rx}
  P.~Nason,
  JHEP {\bf 0411} (2004) 040
  doi:10.1088/1126-6708/2004/11/040
  [hep-ph/0409146].

\bibitem{Frixione:2007vw}
  S.~Frixione, P.~Nason and C.~Oleari,
  JHEP {\bf 0711} (2007) 070
  doi:10.1088/1126-6708/2007/11/070
  [arXiv:0709.2092 [hep-ph]].

\bibitem{Alwall:2014hca}
  J.~Alwall {\it et al.},
  JHEP {\bf 1407} (2014) 079
  doi:10.1007/JHEP07(2014)079
  [arXiv:1405.0301 [hep-ph]].


\bibitem{Khachatryan:2014jba}
  V.~Khachatryan {\it et al.} [CMS Collaboration],
  Eur.\ Phys.\ J.\ C {\bf 75} (2015) no.5,  212
  doi:10.1140/epjc/s10052-015-3351-7
  [arXiv:1412.8662 [hep-ex]].

\bibitem{Sirunyan:2018hoz}
  A.~M.~Sirunyan {\it et al.} [CMS Collaboration],
  Phys.\ Rev.\ Lett.\  {\bf 120} (2018) no.23,  231801
  doi:10.1103/PhysRevLett.120.231801, 10.1130/PhysRevLett.120.231801
  [arXiv:1804.02610 [hep-ex]].

\bibitem{ATLAS:2011tau}
  ATLAS and CMS Collaborations and LHC Higgs Combination Group,
  ATL-PHYS-PUB-2011-011, CMS-NOTE-2011-005.



\end{thebibliography}
\end{document}